\documentclass[journal]{IEEEtran}
\usepackage{arabtex}
\usepackage{utf8}
\usepackage[hyphens]{url}
\usepackage{hyperref}
\usepackage[hyphenbreaks]{breakurl}
\newcommand{\quotes}[1]{``#1''} 
\usepackage{paralist}
\usepackage{sistyle} 
\SIthousandsep{,} 
\usepackage{csquotes} 
\newcommand{\mtnote}[1]{\textsuperscript{\TPTtagStyle{#1}}} 
\newcommand{\autorefappendix}[1]{\hyperref[#1]{appendix~\ref*{#1}}}  
\usepackage{booktabs} 
\usepackage{rotating} 
\usepackage{color,colortbl} 
\definecolor{lightYellow}{cmyk}{0,0.04,0.4,0}
\definecolor{lightPurple}{cmyk}{0,0.27,0,0.07}
\usepackage{xcolor} 
\usepackage{tabularx} 
\newcolumntype{L}{>{\raggedright\arraybackslash}X} 
\usepackage{multicol} 
\usepackage{multirow} 
\usepackage{authblk} 
\usepackage{pdfpages} 
\usepackage{graphicx} 
\usepackage{bmpsize} 
\usepackage{caption} 
\usepackage{textcomp}
\usepackage{placeins} 
\definecolor{gray}{cmyk}{0.57,0.46,0.40,0.25} 
\usepackage{threeparttable} 
\usepackage{amssymb}
\usepackage{pifont}
\newcommand{\cmark}{\ding{51}}%
\newcommand{\xmark}{\ding{55}}%
\usepackage[usestackEOL]{stackengine}

\begin{document}

\title{Saudi Parents' Privacy Concerns about Their Children's Smart Device Applications}

\author[1]{Eman Alashwali}
\author[2]{Fatimah Alashwali}
\affil[1]{King Abdulaziz University, Jeddah, Saudi Arabia \authorcr {\tt ealashwali@kau.edu.sa}}
\affil[2]{Princess Nourah Bint Abdulrahman University, Riyadh, Saudi Arabia \authorcr {\tt fsalashwali@pnu.edu.sa}}

\maketitle

\begin{abstract}
In this paper, we investigate Saudi parents' privacy concerns regarding their children's smart device applications (apps). To this end, we conducted a survey and analysed 119 responses. Our results show that Saudi parents expressed a high level of concern regarding their children's privacy when using smart device apps. However, they expressed higher concerns about apps' content than privacy issues such as apps' requests to access sensitive data. Furthermore, parents' concerns are not in line with most of the children's installed apps, which contain apps inappropriate for their age, require parental guidance, and request access to sensitive data such as location. We also discuss several aspects of Saudi parents' practices and concerns compared to those reported by Western (mainly from the UK) and Chinese parents in previous reports. We found interesting patterns and established new relationships. For example, Saudi and Chinese children have a higher level of autonomy in installing apps on their devices than Western children who mostly request apps installation from their parents. However, more Saudi parents who end up uninstalling apps from their children's devices due to privacy concerns than Western parents. This suggests that parents' involvement in installing children's apps is related to a reduced number of uninstallation incidents, which are mostly negatively received by children. Furthermore, Saudi and Western parents show higher levels of privacy concerns than Chinese parents. Finally, we tested 14 privacy practices and concerns against high versus low socioeconomic classes (parents' education, technical background, and income) to find whether there are significant differences between high and low classes (we denote these differences by \quotes{digital divide}). Out of 42 tests (14 properties $\times$ 3 classes) we found significant differences between high and low classes in 7 tests only. While this is a positive trend overall, it is important to work on bridging these gaps. The results of this paper provide key findings to identify areas of improvement and recommendations, especially for Saudis, which can be used by parents, developers, researchers, regulators, and policy makers.

\end{abstract}

\section{Introduction} \label{sec:intro}
\subsection{Motivation}\label{sec:motivation}
Smart devices such as smartphones and tablets, along with their applications (apps), represent an integral part of almost everyone's daily life. Children are increasingly spending more time using these devices and apps. In the UK, as of 2019, according to research by Ofcom~\cite{ofcom20}, 45\% of children between 5 and 15 years old own a smartphone, compared to 35\% in 2015. In the US, as of 2019, according to the Common Sense census~\cite{commonsense19}, 53\% of children between 8 and 11 years old, and 91\% of children between 8 and 18 years old, own a smartphone, compared to 32\% and 77\% respectively in 2015. In Saudi Arabia, according to a report by \quotes{Norton}~\cite{norton18}\footnote{The report surveyed \num{6986} parents who have children between 5 to 16 years old in 10 different countries from around the globe.}, Saudi Arabia's children rank the highest in accessing\footnote{Owning or borrowing.} tablet devices, where 74\% of them do so. Moreover, Saudi children are the youngest to own their first connected mobile device, where on average, a 7-year-old Saudi child has his/her first connected device~\cite{norton18}. Furthermore, the Internet penetration rate in Saudi Arabia is 93.31\%, according to the 2019 yearbook of the General Authority for Statistics (GAS) in Saudi Arabia~\cite{gas_yearbook_19}.\par 

In recent years, two studies have looked at parents' privacy concerns regarding their children's smart device apps~\cite{zhao18,wang19}. They surveyed Western (mainly from the UK)~\cite{zhao18} and Chinese parents~\cite{wang19}. \par 

In our study, we pay particular attention to Saudi parents' privacy practices and concerns about their children's smart device apps. There are several reasons that make Saudi Arabia an interesting and unique target to study. First, Saudi Arabia is a developing country that was established in 1932 under the name of Kingdom of Saudi Arabia (KSA)~\cite{wikipedia21_sa}, and is located in south-western Asia. Second, it has one of the youngest populations in the world. As of 2019, 31.42\% (\num{10749944}) of Saudi population\footnote{The number of population of Saudi Arabia in mid 2019 according to GAS is \num{34218169}~\cite{gas_yearbook_19}.} are below 20 years old, and 24.52\% (\num{8389963}) are below 15 years old~\cite{gas19}. Third, it is a wealthy country, mainly from oil export revenues which represent 70\% of its export earnings according to the Organization of the Petroleum Exporting Countries (OPEC)~\cite{opec20}. Saudi Arabia has around 17\% of the world's proven oil reserves and is the second largest member of the OPEC~\cite{opec20}. \par 

Saudi Arabia is developing rapidly. Obviously, the strong economy combined with its young population are the two main factors for such rapid development. This in turn resulted in high rates of technology adoption and Internet access. However, parents' awareness about emerging technology issues such as privacy may not be in line with the adoption pace. It is not surprising to hear comments such as the following comments by an orthodontic specialist and a mother of two children, 5 and 8 years old~\cite{gassem19}:
\begin{displayquote}{\textit{I do not know how to use the smart TV anymore because it is more complicated, but the [smart] TV is very simple for my children to use. They know how to switch from YouTube to cable channels and to Netflix. Sometimes I need them to do that for me.}}\end{displayquote}

Internet and mobile app users often share personal data with the website or app operators. These data can be precise location, camera, microphone, or contacts data. Some of these data are necessary for the website or app to function, while others provide optional improvements to the provided service. However, website and app operators might collect data to sell them to third parties. Data monetisation has become a key business model for many websites and apps nowadays. This is a serious concern when that data belongs to children. Users and societies can be harmed if their data was misused by operators. For example, if users' data was shared with third parties without their consent, or if their data was used for purposes they were unaware of. In recent years, a data analytics firm called Cambridge Analytica received a lot of public attention during investigations from the UK's authorities after allegations of misusing millions of Facebook users' data to achieve political goals~\cite{isaak18,cadwalladr18,rosenberg18,confessore18}. Children are often viewed as vulnerable users online~\cite{kennedy19}. They are susceptible to various types of risks associated with data breaches such as cyber bullying~\cite{livingstone14}, cyber stalking, and online sexual harassment. Therefore, children's data in particular, need to be shared with care and parents' awareness of what they are allowing their children to share online.

The need for supporting families in children's online privacy issues has been realised by the Saudi government. Recently, in 2020, the Saudi Crown Prince Muhammad Bin Salman launched an initiative to protect children in the cyber world at the Global Cyber Security Forum in Riyadh, Saudi Arabia~\cite{gazette20}.\par 

Motivated by the aforementioned need, in addition to the lack of comprehensive analytical studies regarding privacy concerns in our region, the main goal of this study is to understand Saudi parents' privacy concerns about their children's smart device apps. The results of this study will help us better understand the status of Saudi families to identify areas of improvement, set recommendations, and develop the right tools towards a safer online world for children. 
\subsection{Scope}
First, in terms of device types, our study's scope is on \quotes{tablet} and \quotes{smartphone} devices. For brevity, we used the term \quotes{smart devices} to refer to both \quotes{tablets} and \quotes{smartphones} in general contexts such as the paper's and survey's titles, abstract, and introductions. However, we used the specific terms \quotes{tablets} and \quotes{smartphones} where clarity is required such as in the survey questions and for the rest of the paper to summarise the results. Second, in terms of children's ages, our scope is on children whose ages are between 6 to 12 years. We chose this range as it represents the ideal primary school age in Saudi Arabia, and possibly many other countries. In addition, the chosen age range represents the pre-teen age.
\subsection{Research Questions}
The main research questions in this study are as follows:
\begin{enumerate}
\item What privacy concerns do Saudi parents have about their children's smart device apps?
\item What are they doing to mediate these concerns?
\item Are there significant differences in privacy practices and concerns about children's smart device apps between high versus low Saudi socioeconomic classes (i.e. is there a \quotes{digital divide}\footnote{The term \quotes{digital divide} is a common term used to describe the gap between the societies that have access to technology and those that do not. While we use it to describe a different sort of gap (i.e. the gap between high versus low socioeconomic classes in terms of practices and concerns about their children's privacy in smart device apps), we find it a suitable term.})? 
\end{enumerate}
\subsection{Contributions}
To the best of our knowledge, our study is the first and largest study to date that comprehensively looked at Saudi parents' privacy concerns about their children's smart device applications. Our results provide a better understanding of Saudis concerns, identify areas of improvement, and guide future efforts. The results can be used by parents, developers, researchers, regulators, and policy makers. To summarise, the main contributions of this paper are: 
\begin{enumerate}
\item We conducted a survey to understand Saudi parents' privacy practices and concerns about their children's smart device apps. We analysed 119 completed responses by Saudi parents. We find that Saudi parents expressed a high level of concern regarding their children's privacy when using smart device apps. However, they expressed higher concerns about apps' content than privacy issues such as apps' requests to access sensitive data. Furthermore, parents' concerns are not in line with most of the children's installed apps, which contain apps inappropriate for their age, require parental guidance, and request access to sensitive data such as location. 

\item We discussed several aspects of Saudi parents' privacy practices and concerns about their children's smart device apps compared to those reported by Western and Chinese parents. We identified interesting patterns and established new relationships, which can be linked to cultural and political differences. 

\item We tested whether there are significant differences between high versus low Saudi socioeconomic classes (parents' education, technical background, and income) in terms of privacy practices (digital divide). Out of 42 tests, we found significant differences between high and low classes in 7 tests only. 
\end{enumerate}
\subsection{Organisation}
This paper is organised as follows: In~\autoref{sec:related}, we summarise related work. In~\autoref{sec:method}, we describe our study methodology. In~\autoref{sec:results}, we discuss our results. In~\autoref{sec:divide}, we examine whether there is a digital divide (significant differences between high versus low Saudi socioeconomic classes). In~\autoref{sec:recomm}, we make recommendations. In~\autoref{sec:limit}, we list the limitations. Finally, in~\autoref{sec:conclusion}, we conclude the paper.

\section{Related Work} \label{sec:related}
In this section, we describe related work which we classify into three categories: 
\begin{inparaenum}
	\item Internet usage and the associated risks,
	\item parents' concerns about children's online safety, and
	\item means to ensure children's online safety.
\end{inparaenum} Our study mainly fits in the second category: parents' concerns about children's online safety. However, we include the other categories for completeness, to provide the full picture of this realm.

\subsection{Internet Usage and the Associated Risks}
While Internet access provides unprecedented opportunities for communication, learning, and entertaining, it brings numerous risks with it. For example, access to global information brings with it the risk of viewing inappropriate or illegal content, access to social networks brings with it the risk of being exposed to bullying and sexual harassment, and the list goes on~\cite{hasebrink09}. In~\cite{hasebrink09}, Hasebrink et al. provided a classification for online risks from two dimensions: the child's role and the provider's motive. \autoref{tab:riskclass} illustrates Hasebrink et al.'s classification.

\begin{table*}[!t]
	\centering
	\caption{Hasebrink et al.'s classification of online risks from two dimensions: the child's role and the provider's motive~\cite{hasebrink09}.}
	\label{tab:riskclass}
	\begin{tabularx}{\linewidth}{lLLLLL}
	\hline 
		\multirow{2}{*}{Child's Role} & & \multicolumn{4}{c}{{Provider's Role} }\\
	    \cline{3-6}
		 & & Commercial  & Aggressive & Sexual   & Values \\
		\hline
		Content: child as recipient& & Advertising, exploitation of personal information  & Violent web content & Problematic sexual web content & Biased information, racism, blasphemy, health \quotes{advice} \\
		\hline
		Contact: child as participant & & More sophisticated exploitation, children being tracked by advertising  & Being harassed, stalked, bullied  & Being groomed, arranging for offline contacts & Being supplied with misinformation \\ 
		\hline
		Conduct: child as actor &  & Illegal downloads, sending offensive messages to peers & Cyberbullying someone else, happy slapping  & Publishing porn  & Providing misinformation \\
		\hline
	\end{tabularx}
\end{table*} 


\subsection{Parents' Concerns about Children's Online Safety}
Zhao et al.~\cite{zhao18} surveyed 221 parents with children aged 6 to 10 years old from Western countries (mainly the UK, where 78\% of the respondents were from the UK) regarding what privacy concerns parents have about their children's mobile apps. They identified patterns for practices and concerns. For example, they found that when choosing an app for their children, parents care about apps' content more than the personal data that the apps might collect. Additionally, they found that some of the most widely used apps by children are inappropriate for their age. Subsequently, Wang et al.~\cite{wang19} conducted a related study that surveyed 593 parents with children from 6 to 10 years old from China. Wang et al. compared some of their findings with those reported by Zhao et al.~\cite{zhao18} on Western parents. Interestingly, Wang et al. found some different trends between the two populations, possibly due to political and cultural reasons. For example, Chinese children showed higher levels of autonomy in installing apps in their devices than Western children. Additionally, Western parents showed higher levels of privacy concerns than their Chinese counterparts. Several other studies conducted surveys to investigate certain aspects regarding children's online safety. For example, Dedkova et al.~\cite{dedkova20} analysed 331 responses from Czech parents of children between 5 to 17 years old, to investigate parents' sources for digital security advice, and the relationship between parents' preferred sources and their personal characteristics such as parents' technical skills. They identified four groups of parents based on the sources they used for digital security advice. Moreover, they found that parents' preference for certain sources is related to their Internet skills. Another survey conducted by the C.S. Mott Children's Hospital~\cite{regents21}, asked parents of children from 7 to 12 years old about their children's usage of social media apps, how parents choose safe apps for their children, and parents' use of parental control apps. Among their findings, they found that 17\% of parents whose children use social media are not using any parental controls. Parents in the study also reported the challenges they face in monitoring their children's usage of social media such as the lack of information needed to set up parental control (21\%), and the children's abilities to circumvent the parental controls (32\%). Other studies investigated children's use of mobile devices at home and investigated parents' socioeconomic and educational backgrounds in relation to children's mobile usage~\cite{papadakis19,papadakis22}. Although~\cite{papadakis19,papadakis22} are not investigating security and privacy aspects, they are related to our work as they attempted to identify relationships between parents' socioeconomic and educational backgrounds in relation to their children's mobile usage at home. For example, in~\cite{papadakis19} Papadaki et al. surveyed 293 Greek families and examined parents' socioeconomic and educational backgrounds in relation to children's mobile usage in Greek children~\cite{papadakis19}. They found a relationship between parents' socioeconomic and educational backgrounds with awareness about mobile learning technologies. For example, lower educated or older parents struggle to adapt to the rapid technological advancement~\cite{papadakis19}. We aim to extend this research line to study Middle Eastern parents by surveying Saudi families. \par 

With respect to Saudi families, there are few studies that have investigated children's online safety from various perspectives. In~\cite{alqahtani16}, A. M. Alqahtani surveyed 115 adolescents from the Gulf Cooperation Council (GCC) countries including, Saudi Arabia, who were studying in the UK, and their ages ranged between 13 and 18 years. The survey consisted of several high-level questions about keeping safe online. The study found that online risks are not fully understood by adolescents or their parents. The results also suggest that GCC adolescents' online behaviour differs from their UK counterparts in terms of Internet usage, which may be a route that GCC adolescents take in future. In~\cite{alqahtani17}, N. Alqahtani conducted a study on 30 Saudi parents and their children (30 children) to investigate associations and correlations between what children say about online risks and what their parents do to mediate those risks. They found a disparity between parents' perspectives and their children's online behaviour. In addition to that, they found a lack of collaboration between parents and children to provide online safety. In~\cite{almogbel15}, Almogbel et al. examined the relationship between parents' educational and economical levels against parental control over their children's Internet usage. The study concluded that parents' demographic variables appear to have no effect on the level of control for their children's browsing behaviour. Unlike previous work in the Saudi Arabia context, our study is focused on mobile apps and privacy concerns, which none of the previous work investigated. Furthermore, while Almogbel et al. provided a high-level examination for the relationship between parents' socioeconomic status and their children's Internet usage in~\cite{almogbel15}, our study provides a new perspective. We provide an in-depth examination of a different and wider set of parameters (14 properties about parents' practices and concerns against 3 socioeconomic aspects) in a larger population, to identify whether there are significant differences between high vs. low Saudi socioeconomic classes in terms of privacy practices and concerns. \par


\subsection{Means to Provide Children's Online Safety}
Hartikainen et al.~\cite{hartikainen16} classified the means for ensuring children's online safety as follows: 
\begin{enumerate} 
	\item Industry mediation: such as age restriction, reporting mechanisms, and automatic filtering techniques.  
	\item Policies and educational efforts: such as digital literacy programs, teaching online safety in schools, and creating guidebooks for children and parents.   
	\item Social mediation: such as conversations with children about Internet usage, limiting, or monitoring it.
	\item Technical mediation: such as installing malware detection tools, or parental control software.
	\item Developmental process: such as viewing children's online safety as a developmental process of youth growth.
\end{enumerate} 

However, most of the aforementioned mediation means have limitations. For example, industry mediation such as reporting tools have their technical issues including reliance on users or administrators, e.g. to report abuse. In addition, automated screening tools often suffer from false-positive alerts. Policies, educational, and social mediation such as conversations are not trivial. For example, children can be more competent in emerging technologies than their parents and teachers. Therefore, parents and teachers may be unfamiliar and struggle with emerging technologies. Technical mediation has technical limitations that affect its efficiency. For example, some parental control software can be circumvented by children. In addition, some might highlight ethical concerns regarding children's privacy when parents use parental control software to monitor them. \par   

Designing the right tools for children's online safety is non-trivial. Finding the right balance between technical efficiency, transparency, and children's privacy is challenging. Ghosh et al. analysed reviews of parental control apps posted by children~\cite{ghosh18}. They found that children's reviews are significantly lower than parents' reviews (76\% of the children's reviews are one star)~\cite{ghosh18}. Clearly, children are not viewing these tools positively. However, there might be profound reasons behind this phenomena that needs further research.

\section{Methodology} \label{sec:method}
In this section, we describe our methodology. We start by describing our targeted sample, followed by how we designed and deployed our survey.

\subsection{Targeted Sample}\label{sec:sample}
Our targeted sample was those who identified themselves as: \begin{inparaenum}[1)] 
\item living in Saudi Arabia (either as citizens or residents\footnote{By residents, we refer to non-Saudis living in Saudi Arabia. However, as we stated in the results~(\autoref{sec:responses}), this report only includes responses from Saudi citizens who live in Saudi Arabia, because we wanted to focus on Saudi parents' practices and concerns.}), and 
\item have one or more children whose ages are between 6 to 12 years who use smart device apps on a regular basis\end{inparaenum}.
We stated the aforementioned two conditions in the survey invitation letter. However, to ensure correctness, we included several questions in the survey that enabled us to filter out unwanted responses (i.e. those that did not meet the aforementioned conditions). For example, the survey included a question about whether the respondent is living in Saudi Arabia or elsewhere. This question enabled us to filter out responses coming from outside Saudi Arabia. Furthermore, to check the respondent child's age, we included a multiple-choice question about the child's age with choices ranging from 6 to 12 years. Therefore, a respondent who does not have a child within the specified age range would not find the correct answer, and ideally should not participate in the survey.

\subsection{Survey Design}\label{sec:design}
Our survey consisted of 51 questions divided into the following sections: \begin{inparaenum}[1)] \item introduction, \item demographics and device usage, \item parents' privacy controls and concerns, and finally \item reflection\end{inparaenum}. The majority of questions were multiple choice questions, while several questions required short textual inputs. We used the \quotes{logic} feature in designing our survey. That is, some questions appeared only if certain conditions in the previous question(s) were met. For example, a question appeared only if the previous question was answered with \quotes{yes}. Therefore, the actual number of questions that each participant received varied based on their answers. Our survey was delivered in the Arabic language, the native language of Saudi Arabia. We chose to use Arabic because we wanted to reach a wide range of demographics in the Saudi society, including non-English speakers. Our survey was based on Zhao et al's survey in~\cite{zhao18}\footnote{We received it by email communication with Jun Zhao, the first author of~\cite{zhao18}.}. We modified Zhao et al's survey~\cite{zhao18} in terms of both breadth (more questions added) and depth (several questions and answers adjusted and improved). For example, we added questions to ask about the family income because we wanted to analyse socioeconomic aspects. We also modified some questions such as those asking parents to rate frequency, where we changed the subjective frequency scale used in~\cite{zhao18} (e.g. \quotes{very often}, \quotes{sometimes}, \quotes{occasionally}, ...) to an objective scale (e.g. \quotes{at least once a day}, \quotes{at least once a week}, \quotes{at least once a month}, ...) because we wanted to have quantified objective answers. However, many questions in Zhao et al.'s~\cite{zhao18} survey remained common with our survey, which allowed us to compare several aspects of both studies. We translated Zhao et al's survey~\cite{zhao18} questions mainly using the Google Translate\footnote{https://translate.google.com}, followed by manual editing as needed. The translation was straightforward as the authors are both native Arabic speakers and proficient in English. Moreover, the translated survey was tested on multiple participants who are native Arabic speakers, during the pilot deployment stage (see~\autoref{sec:deployment_pilot} for details about the pilot survey). We designed the survey in an electronic format using the \quotes{alchemer}\footnote{\url{https://alchemer.com}, formerly known as \url{https://surveygizmo.com}}, an online survey platform that enables an advanced level of customisation, and supports Arabic. It also provides several advanced features including reports and customised queries of the results. We used a professional license that provides advanced survey features such as the \quotes{logic} feature which we used in the survey questions. \par
 
The survey started with a login page (not counted as a survey section), which asked participants to enter the survey password which was enclosed within the survey invitation. The password meant to protect our survey since it was publicly available online. The first section of the survey is \quotes{Introduction}, which contains all the information that the participants need to know before starting the survey such as the purpose of the study and the \quotes{informed consent}. The second section is \quotes{Demographics and Device Usage}, which asks about the parent(s) and their child's demographic data, in addition to the child's device usage. The third section is \quotes{Parents Privacy Controls and Concerns}, which is the core of the survey. It consists of four subsections: \begin{inparaenum}[1)] \item \quotes{Installing Apps for Your Child}, \item \quotes{General Practice}, \item \quotes{Privacy Concerns}, and \item \quotes{Parental Controls}. \end{inparaenum} Finally, the fourth section of the survey is \quotes{Reflection}, which contains one optional open question, to enable the participants to reflect on what they have learned from the survey, or to provide further comments, if any. \par    

Participation in the survey was voluntary. We did not offer reimbursement of any kind to respondents to avoid multiple responses, or inaccurate responses, from irresponsible individuals who might participate only for the sake of reimbursement, without fulfilling the survey conditions. The full survey we distributed is available upon request from the corresponding author.

\subsection{Ethical Considerations}\label{sec:ethical}
We followed the common ethical guidelines required for studies that involve human subjects. We obtained a formal ethical approval for our study from our institution's (King Abdulaziz University (KAU)) research ethics committee. Participants were provided with an introduction page. In it, we informed them about the research title and goals, how their data will be collected, processed, stored, secured, and for how long, in addition to all the information they needed to know before taking part in the survey. Participants were asked to click the \quotes{Next} button (we used an online survey) if they agree (consent) to take part. They were also informed that they can quit the survey at any point without any consequences, by closing the browser. Furthermore, we did not collect personally identifiable data (e.g. name, national ID, and work place). Participants were informed that the results of this study will be published in an anonymous format. Finally, participants were provided with our contact details, if they had any concerns.

\subsection{Survey Deployment}\label{sec:deployment}
We deployed the survey in two phases: a pilot survey, followed by the actual survey. In what follows, we describe each phase.

\subsubsection{Pilot Survey} \label{sec:deployment_pilot}
Before we deployed the survey, we invited three participants from our personal network who were willing to participate and provide us with feedback through a short discussion after they finish the survey. In terms of their demographics, they come from diverse socioeconomic and educational levels. All of them are females. Two of those participants satisfied the survey conditions: Saudis and have one or more children who are between 6 to 12 years old who use smart device apps on a regular basis. The third participant is Saudi but did not have children. Therefore, we asked her to answer the child-related questions based on an imaginary child that she built in her mind. Since the pilot survey responses would not be included in the results, the correctness of the participants' answers was not important. Our goal from the pilot survey was to gather overall feedback, test the online survey deployment from different device types and manufacturers (we ensured that our pilot participants use different types of devices), and measure the time needed to complete the survey. We asked the participants to write down their comments, if they had any. After they finished the survey, we had a short conversation with each participant. We identified several issues reported by them. We modified the survey based on the pilot respondents' comments, if the comment made sense to us too. For example, we received several comments regarding the survey's length (that it was long). Hence, to reduce the survey's length, we removed several questions at the end of the survey that were unnecessary to our study. For example, we removed a question that asks whether the participant is willing to take part in an interview in the future, as we were not certain whether we would need to conduct interviews in the future. Moreover, this question is unhelpful without taking the participant's contact details, which might reveal the participant's identity, while we wanted to keep our survey anonymous. We also removed a question that asks whether the participant would go through some privacy settings with his/her child after the study. Although this question might add further insights to our analysis, it was unnecessary, and we gave higher priority to reduce the survey length. We also received a comment regarding using the term \quotes{partner}, which may be incompatible with our conservative society. Hence, we replaced it with the term \quotes{spouse}. We also fixed several typos. We then communicated the changes to the respondents and considered any further comments. We repeated this process until the participants had no further valid comments on the survey. This process took multiple rounds to be completed. 

\subsubsection{Actual Survey} \label{sec:deployment_actual}
Once we reached a satisfactory version of the survey after the pilot survey phase, we distributed the survey through an invitation letter which contained a link (URL)\footnote{Using the secure HTTPS protocol.} to our online survey. We sent the invitation either individually or through groups (mainly WhatsApp\footnote{WhatsApp (\url{https://www.whatsapp.com}), a messaging app.} groups), to our personal and professional networks. We targeted groups and individuals from both genders males and females, from different socioeconomic and educational backgrounds, to reach a diverse sample of participants. The survey was distributed mainly using WhatsApp, email, and LinkedIn\footnote{LinkedIn (\url{https://linkedin.com}) is a professional networking online platform.}. For the WhatsApp groups, they were diverse: family, personal, and professional networking groups. Combined together, they contained members from diverse demographics. We posted the LinkedIn invitation through the first author's LinkedIn account, where the invitation was visible to her professional network connections only. In addition, we distributed the survey to our university's (KAU) staff members (both administrative and academic staff who have diverse socioeconomic and educational backgrounds) through a mailing list through the university's (KAU) Deanship of Postgraduate Studies. Our university (KAU) is a large state university in the western region of Saudi Arabia. It has over 4000 academics, and 4000 administrative staff members~\cite{qs21,wikipedia20}. In the invitations to our personal and professional networks, we added a statement to encourage the recipients to circulate them among their groups, especially groups that have members from diverse social and educational backgrounds, and to parents' groups of their children's schools. Parents' WhatsApp groups are common among parents in our society. We avoided sharing the survey link over \quotes{public} social media accounts which may contain random followers whom we neither know directly nor have a common connection with. This is to avoid incorrect or duplicate responses from irresponsible users of such public social networks, or possibly from automated software bots. Having said that, our invitation was not limited to the authors' networks only, which are already diverse and large enough. As stated earlier, we encouraged the recipients to circulate the invitation among their groups and networks, including parents' groups of their children's schools. The survey was open to participants for 10 days, from November 16 to November 25, 2020.

\subsection{Data Analysis}\label{sec:analysis}
This section is meant to provide a brief overview of the data analysis method we used. More details are embedded in the rest of the paper as needed. After we closed the survey, we gathered the data for analysis. We employed both descriptive analysis and exploratory analysis to summarise results, explore data, and find new relationships~\cite{educba20}. For the comparisons with the two previous reports~\cite{zhao18,wang19}, we conducted high-level comparisons (a form of discussion) between our results and~\cite{zhao18,wang19} results in some common questions and aspects, to identify patterns or relationships, if any. The previous two studies~\cite{zhao18,wang19} were conducted independently of ours. Hence, there are variations between the three studies in terms of demographics and participants' recruitment. In~\cite{zhao18,wang19}, the children's age ranged from 6 to 10, while in our study they ranged from 6 to 12. However, the children samples of the three studies can be grouped into primary school and pre-teen children age, hence the findings are comparable from this aspect. In terms of participants' recruitment, both~\cite{zhao18,wang19}, and our study are based on online surveys. However, each study distributed the survey link using a different method. In~\cite{zhao18}, they used Prolific\footnote{https://www.prolific.co}, a crowd sourced online survey platform,~\cite{wang19} recruited participants through local primary schools in the Shanghai area, while we recruited participants by circulating the invitation through several channels such as WhatsApp groups and a large university mailing list, which contained members from both genders, and diverse socioeconomic and educational backgrounds (see~\autoref{sec:deployment_actual} for further details on the deployment). It should be noted that our comparisons with~\cite{zhao18,wang19} are not meant to be direct comparisons. They meant to be a form of discussions to identify patterns and new relationships, if any. Therefore, we accepted the variations in demographics and recruitment methods between the studies, given that they were conducted independently of ours, and that our study is non-interventional. That is we are not measuring an effect of a specific intervention. Rather, we compare and discuss patterns of practices and concerns. Finally, in~\autoref{sec:divide}, to identify whether there is a digital divide between high versus low Saudi socioeconomic classes, we used the Chi-square test of independence or the Fisher-Irwin test (the latter used only if the expected frequency is less than 5)~\cite{campbell07}.

\section{Results}\label{sec:results}
\subsection{Responses} \label{sec:responses}
We received 136 completed responses. In this paper, we excluded a total of 17 responses for the following reasons: first, we excluded 3 responses for respondents who reported that they are living outside Saudi Arabia, since our study is for those who live in Saudi Arabia. Second, we also excluded 2 responses from participants who described their relationship with the child as an \quotes{aunt} and as a \quotes{grandmother} as we decided to count responses from those who have a formal guardianship with the child (either a \quotes{parent} or a \quotes{guardian}), to avoid inaccurate responses. Third, while we were initially interested in understanding both Saudi citizens and residents (non-Saudis living in Saudi Arabia), in this report, we decided to focus only on Saudi citizens. As a result, we excluded 12 responses from those who identified themselves as non-Saudis. \par 

We end up with 119 completed responses from respondents who identified themselves as: \begin{inparaenum}[1)] \item Saudi citizens, \item living in Saudi Arabia, and \item parents or guardians of at least one child whose age is between 6 to 12 years who use smart device apps in a regular basis\end{inparaenum}. In what follows, we report the survey results in detail.

\subsection{Demographics and Device Usage} \label{sec:demographics}
\subsubsection{Parents' Demographics} \label{sec:demo}
In terms of geographic regions, as listed in~\autoref{tab:parents_demo}, our respondents are mainly from the western region of Saudi Arabia (73.95\%), followed by the central region (16.81\%), while the remaining minorities are from the Eastern, Northern, and Southern regions of Saudi Arabia. The regional bias is very likely due to the fact that the first author (principal investigator) lives in the western region of Saudi Arabia (Jeddah city). However, the Western region (a.k.a. Makkah Region) is a very large and diverse region, containing several cities with a population of \num{9033491} in mid 2019~\cite{gas_yearbook_19}. In terms of parents' gender, most of the respondents are females (86.55\%), despite the fact that we distributed the survey to groups that contain both males and females. It is worth noting that a similar gender bias towards female respondents has been observed in two similar studies, one in the UK~\cite{zhao18}, and the other in China~\cite{wang19}.

\begin{table}[!t]
	\centering
	\caption{Parents' general demographics.}
	\label{tab:parents_demo}
	\begin{tabular}{ll@{\hspace{5pt}}r}
		\toprule
		\multicolumn{3}{c}{\textit{N = 119}} \\
		\hline 
		Region			 & \multicolumn{2}{c}{Participants} 			\\
		\hline
		\quad Western 	 & 88 & (73.95\%) \\
		\quad Eastern    & 6  & (5.04\%)  \\
		\quad Central    & 20 & (16.81\%) \\
		\quad Northern   & 3  & (2.52\%)   \\
		\quad Southern   & 2  & (1.68\%)   \\ 
		\hline 
		Age				& & \\
		\hline
		\quad 18 to 24		& 0  & (0\%) \\
		\quad 25 to 34		& 28 & (23.53\%) \\
		\quad 35 to 44		& 66 & (55.46\%) \\
		\quad 45 to 54 		& 22 & (18.49\%) \\
		\quad 55 to 64 		& 3 & (2.52\%) \\
		\quad 65 or above	& 0  & (0\%) \\
		\hline 
		Gender & &	\\
		\hline
		\quad Female 		& 103 & (86.55\%) \\
		\quad Male 			& 16  & (13.45\%) \\
		\hline 
		Number of Children not older than 18 years old & & \\
		\hline 
		\quad 1 			& 8	  & (6.72\%)  \\
		\quad 2				& 49  & (41.18\%)   \\
		\quad 3 			& 37  & (31.09\%)   \\
		\quad 4				& 20  & (16.81\%)  \\
		\quad 5				& 3   & (2.52\%)  \\
		\quad 6 			& 1	  & (0.84\%)   \\
		\quad 7 			& 1	  & (0.84\%)  \\
		\quad 8 			& 0	  & (0\%)  \\
		\quad 9 			& 0	  & (0\%)  \\
		\quad 10 			& 0	  & (0\%)  \\
		\quad 11 or more 	& 0	  & (0\%)  \\
		\hline
		Monthly family income in SAR & & \\
		\hline 
		Less than \num{6000} 				  & 8 & (6.72\%) \\
		From \num{6000} to \num{13000} 		  & 22 & (18.49\%) \\
		More than  \num{13000} to \num{20000} & 24 & (20.17\%) \\
		More than \num{20000} to \num{27000}  & 19 & (15.97\%) \\
		More than \num{27000} to \num{34000}  & 12 & (10.08\%) \\
		More than \num{34000} to \num{41000}  & 7 & (5.88\%) \\
		More than \num{41000} to \num{48000}  & 7 &  (5.88\%) \\
		More than  \num{48000}				  & 12 & (10.08\%) \\
		Prefer not to answer 				  & 8 &  (6.72\%) \\
		\bottomrule
	\end{tabular}
\end{table}

In terms of age, most of our respondents are middle-aged, where 55.46\% range from 35 to 44 years old, and 23.53\% are from 25 to 34 years old. The average number of children (who are not older than 18 years old) in the respondents' families is \num{2.73} children. The participants' relationship with the child selected for the survey is mostly a parental relationship as 98.32\% reported, while the remaining had a guardianship relationship. For brevity, in this paper, we use the term \quotes{parent(s)} to refer to both types of respondents: parents and guardians. See~\autoref{tab:parents_demo} for further details about parents' demographics, and~\autoref{tab:child_demo} for details about children's demographics.


\subsubsection{Parents' Economical Status} \label{sec:economic}
Our respondents come from various economical statuses. We classified participants' household monthly income into several classes. Our income classification is mainly based on the \quotes{sufficiency line} (i.e. the bare minimum line), a term coined by Aldamegh in~\cite{aldamegh14}, who conducted a study that sets the minimum required monthly income for a 5 person family to have a decent standard of living in Saudi Arabia, which is meant to describe a slightly better line than the commonly used \quotes{poverty line}. The sufficiency line income varies between \num{6037.57} and \num{12673} Saudi Arabian Riyal (SAR), according to Aldamegh in~\cite{aldamegh14} (depending on the region of Saudi Arabia). However, on average (i.e. for all regions), the sufficiency line income in Saudi Arabia is \num{8926.10} SAR, according to Aldamegh in~\cite{aldamegh14}. Since Aldamegh's study is based on data from 2013, we considered the inflation rate in Saudi Arabia from 2013 to 2020 according to data from~\cite{worlddata2020} which we cross-checked with data from~\cite{statista21}. Accordingly, we updated the sufficiency line monthly income range to be between \num{6342.65} and \num{13313.38} SAR, and the average sufficiency line monthly income to become \num{9377.14} SAR.\par

From the sufficiency line, we defined the monthly household income. We calculated further economical levels using the same approximate interval that appears in the sufficiency line range ($\sim$\num{7000} SAR). Our monthly household classifications are as follows: 
\begin{enumerate}
\item We denote the monthly household income that is less than \num{6000} SAR as \quotes{low income}.
\item We denote the monthly household income that ranges from \num{6000} to \num{13000} SAR as \quotes{sufficiency line income}.
\item We denote the monthly household income that ranges from more than \num{13000} to \num{20000} SAR as \quotes{low-middle income}. 
\item We denote the monthly household income that ranges from more than \num{20000} to \num{27000} SAR as \quotes{middle income}. 
\item We denote the monthly household income that ranges from more than \num{27000} to \num{34000} SAR as \quotes{high-middle income}. 
\item We denote the monthly household income that ranges from more than \num{34000} and upwards as \quotes{above-middle income}.
\end{enumerate}

From there, there can be several unspecified further levels of the \quotes{above-middle income}. In addition to the sufficiency line base that we used from Aldamegh~\cite{aldamegh14} to build our classification, we also checked the 2018's GAS average household income for Saudi citizens~\cite{gas18b}. Overall, in Saudi Arabia, the average household income for Saudis is \num{13603} SAR, and in the Western region (where most of our respondents are from) is \num{14648}, and in the capital city Riyadh is \num{16011}~\cite{gas18b}. Based on the aforementioned data, our definitions of low income, low-middle, middle, high-middle, and above-middle income classes are sound. \par

The last block of~\autoref{tab:parents_demo} shows the economic status levels that we defined along with the responses we received, which are sufficiently diverse. Those in the middle income class (including \quotes{low-middle}, \quotes{middle}, \quotes{high-middle}) ranges represent a total of 46.22\% of the respondents, while 18.49\% are in the sufficiency line, 6.72\% are in the \quotes{low income}, 21.85\% are in the \quotes{above-middle income}, and 6.72\% of the respondents preferred not to answer this question.


\subsubsection{Parents' Employment Status}
As demonstrated in~\autoref{tab:parents_edu}, our participants have various employment statuses. The majority of the respondents (64.71\%), and their spouses (74.79\%), are full-time paid employees. We have a high percentage of respondents (18.49\%) who identified themselves as \quotes{full-time housewife/househusband}\footnote{One participant chose \quotes{Other} in the employment status question and specified her choice as \quotes{housewife}. Therefore, the actual number of full-time housewives is 23 (19.33\%).}. However, all of them are females. This is not so surprising since financial independence for women in Saudi Arabia is non-obligatory. That is, men are obliged (by the Islamic \quotes{Shariaa} law) to finance women in their families who are under their guardianship (responsibility), e.g. a husband is obliged to finance his wife and unmarried daughters. Nevertheless, the majority of women choose to work for various reasons including financial reasons such as having their financial independence and to support their families financially.

\begin{table}[!t]
	\centering
	\caption{Parents and their spouses' educational, employment, and technical backgrounds. Columns with \quotes{-} value mean we do not have figures for them since participants were not offered \quotes{Don't know}, \quotes{N.A.} answers for questions about themselves, but were offered them for questions about their spouses.}
	\label{tab:parents_edu}
	\begin{tabularx}{\columnwidth}{ll@{\hspace{5pt}}rl@{\hspace{5pt}}r}
		\toprule
		\multicolumn{5}{c}{\textit{N = 119}} \\
		\hline 
		Highest degree  &\multicolumn{2}{l}{Participants} & \multicolumn{2}{l}{Spouses}  \\
		\hline 		
		\quad Doctorate				   & 28 & (23.53\%) &	13	&	(10.92\%)	\\
		\quad Master's			   & 31 & (26.05\%) &	22	&	(18.49\%)	\\
		\quad Bachelor's			   & 48 & (40.34\%) &	55	&	(46.22\%)	\\
		\quad Diploma 			   & 1  & (0.84\%)  &	6	&	(5.04\%)	\\
		\quad High School		   & 8  & (6.72\%)  &	15	&	(12.61\%)	\\
		\quad Intermediate school  & 2  & (1.68\%)  &	2	&	(1.68\%)	\\
		\quad Primary school	   & 1  & (0.84\%)  &	1	&	(0.84\%)	\\
		\quad Don't know		   & -	& -      & 0	& (0\%) \\
		\quad N.A.				   & -  & - 	 &  5	&   (4.20\%)	\\
		\quad Other				   & 0  & (0\%)    & 0    & (0\%)  \\
		\hline
		Employment status  		   &  & & & \\
		\hline
		\quad Full-time paid employee & 77 & (64.71\%) & 89 & (74.79)\% \\
		\quad Full-time housewife/househusband  & 22 & (18.49\%) & 10 & (8.40)\% \\
		\quad Part-time paid employee & 1 & (0.84\%) & 2  & (1.68\%) \\
		\quad Self-employed or business owner & 4 & (3.36\%) & 7 & (5.88\%) \\
		\quad Unemployed, looking for work & 6 & (5.04\%) & 2 & (1.68\%) \\
		\quad Unemployed, not looking for work & 3 & (2.52\%) & 0 & (0\%) \\
		\quad Unable to work				   & 0 & (0\%) & 0 & (0\%) \\
		\quad Student & 4 & (3.36\%) & 1 & (0.84\%) \\
		\quad Retired & 1 & (0.84\%) & 6 & (5.04\%) \\
		\quad Don't know		   & -	& -      & 0	& (0\%) \\
		\quad N.A.   & - & -     & 2 & (1.68\%) \\
		\quad Other	  & 1 & (0.84\%)   & 0 & (0\%) \\
		\hline
		Technical background  		   &  & & & \\
		\hline
		\quad Yes					   & 35 & (29.41\%) & 21 & (17.65\%) \\
		\quad No					   & 84 & (70.59\%) & 	88 & (73.95\%) \\
		\quad Don't know			   & -  & -	 	 & 0   & (0\%) \\
		\quad N.A.					   & -  & -	 	 & 10  & (8.40\%) \\
		\bottomrule
	\end{tabularx}
\end{table}

\subsubsection{Parents' Educational, and Technical Backgrounds}
Our respondents come from various educational and technical backgrounds. In terms of educational background, the majority have a university degree, where 40.34\% have Bachelor's, 26.05\% have Master's, and 23.53\% have Doctorate degrees. Despite our efforts to distribute our survey to parents with various educational backgrounds, our respondents are biased towards highly educated parents. A national survey conducted in 2017 by GAS~\cite{gas_edu_17} showed that, of the adult population whose ages were 25 years or more\footnote{We extracted the national education figures for adult population aged 25 years or more to make the national education survey figures comparable to our survey figures since all our respondents are aged 25 years or more.}, 0.35\%  have Doctorate, 1.12\% have Master's, and 25.03\% have Bachelor's degrees. In terms of technical background, we defined this as having a degree or working in Computer Science (CS), Information Technology (IT), Information Systems (IS), or Computer Engineering (CE) fields. We communicated this definition to our participants within the concerned question in the survey.~\autoref{tab:parents_edu} shows the parents and their spouses who have a technical background. There are 16 (13.45\%) of the responses where both parents have technical background.~\autoref{tab:parents_edu} shows the parents' employment, educational, and technical backgrounds in more detail.

\subsubsection{Children's Demographics}
The average age of the child selected for the survey is \num{9.14} years old. The children's gender is almost equally distributed, where 50.42\% are females, and 49.58\% are males. This is nearly equivalent to the gender distribution of the Saudi populations' census for children between 5 to 14 years old (49.09\% females, and 50.91\% males), according to the 2019 GAS year book~\cite{gas_yearbook_19}. In terms of the child's school type, 66.39\% are in private schools (either International or non-International), and 33.61\% are in government (state) schools\footnote{One participant chose the school type as \quotes{Other} and specified it as a \quotes{talented school}, which is a type of government schools for talented students. Therefore, we added this participant's answer to the percentage of the government schools in our summary (but we left the figures intact in~\autoref{tab:child_demo}).}. Since private schools are with fees, and state schools are free, the school type can indicate the family's economic status to some extent\footnote{Some exceptions can occur with scholarships. However, scholarships are very uncommon at the schools level in Saudi Arabia.}.~\autoref{tab:child_demo} shows the children's demographics in more detail.

\begin{table}[!t] 
\centering
\caption{Children's demographics.}
\label{tab:child_demo}
\begin{tabular}{ll@{\hspace{5pt}}r}
\toprule
\multicolumn{3}{c}{\textit{N = 119}} \\
\hline
Age & \multicolumn{2}{c}{Participants}\\	
\hline 				
\quad 6 				& 17 & (14.29\%) \\
\quad 7					& 19 & (15.97\%) \\	
\quad 8 				& 11 & (9.24\%) \\	
\quad 9 				& 16 & (13.45\%) \\		
\quad 10 				& 18 & (15.13\%) \\		
\quad 11 				& 15 & (12.61\%) \\		
\quad 12 				& 23 & (19.33\%) \\					
\hline 
Gender & & \\
\hline 
\quad Female		    & 60 & (50.42\%) \\
\quad Male				& 59 & (49.58\%) \\
\hline 
Parental Relationship & & \\
\hline 
\quad Parent			& 117 & (98.32\%) \\
\quad Guardian			& 2   & (1.68\%) \\
\hline
Child's school type	& & \\
\hline 	
\quad Private (International)     & 42    & (35.29\%) \\
\quad Private (non-International) & 37	  & (31.09\%) \\
\quad Government 				  & 39	  & (32.77\%) \\
\quad Other						  & 1 	  & (0.84\%) \\
\hline
Internet access	(screen time) & & \\	
\hline 
\quad No access at all 	& 3 & (2.52\%) \\
\quad Less than 2 hours & 16 & (13.45\%) \\
\quad Between 2 to 4 hours & 41 & (34.45\%) \\
\quad More than 4 hours & 59 & (49.58\%) \\
\hline 
Device ownership & & \\
\hline 
\quad Has his/her separate device & 82 & (68.91\%) \\
\quad Shares one device with his/her sibling(s) & 14 & (11.76\%) \\	
\quad Shares  one device with one of his/her parents & 22 & (18.49\%) \\
\quad Other & 1 & (0.84\%) \\
\bottomrule
\end{tabular}
\end{table}
		
\subsubsection{Children's Smart Device Usage}
\begin{figure} 
\centering
\includegraphics[width=\columnwidth]{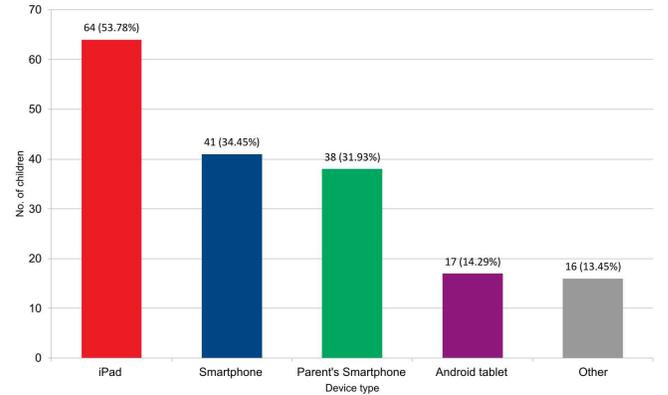}
\caption{Devices owned by children.}
\label{fig:devices}
\end{figure}

With respect to children's used devices, \autoref{fig:devices} shows the children's most commonly used devices. Note that a child might have access to more than one device and we allowed parents to select all that apply. iPads are the most widely used devices by children (53.78\%), followed by smartphones (34.45\%), a parent's smartphone (31.93\%), and Android tablet (14.29\%). The \quotes{Other} category is selected by 13.45\% of the participants who listed devices such as Desktop, Laptop, Macbook, Nintendo, Smart TV, and Play Station. \par 

The majority of children have their own devices (68.91\%), while the rest share the device with their siblings (11.76\%), or one of their parents (18.49\%), and one (0.84\%) chose \quotes{Other} and specified that the child is sharing the device with both parents and siblings. While sharing a parent device with his/her child might give the parent perceived control over the child's usage of the device, it should be shared with caution. Parents' smart devices may contain apps with inappropriate content for the child's age, or apps that contain confidential information such as confidential emails in the parents' work email app.\par 

In terms of the time children spend on the Internet (a.k.a. \quotes{screen time}), we classified the daily time children spend on the Internet into 3 categories: \quotes{very high} (more than 4 hours per day), \quotes{high} (between 2 to 4 hours per day), \quotes{low} (less than 2 hours per day), in addition to \quotes{no access at all}. There are various hourly thresholds used to report screen time. However, we based our classification on a scale from~\cite{thomas19,sandercock13}. The American Academy of Child and Adolescent Psychiatry (ACAP) recommends limiting screen time for children whose ages range between 2 to 5 years to around 1 hour per weekday, and up to 3 hours during weekend days. Moreover, ACAP recommends limiting (without specifying the limit) screen time for children starting from 6 years~\cite{aacp21}. Limiting screen time for children is a controversial issue: some studies recommend no more than 2 hours per day and link high screen time to negative effects such as obesity~\cite{barlow07}, while others argue that there is little evidence that supports implementing screen time limits for children~\cite{przybylski17}.\par 

In our study, parents' responses show that, nearly half of the children (49.58\%) spend \quotes{very high} screen time on tablets and smartphones, 34.45\% spend \quotes{high} screen time, 13.45\% spend \quotes{low} screen time, while 2.52\% do not spend time on the Internet at all.~\autoref{tab:child_demo} demonstrates the results.

\subsection{Apps' Privacy Practices} \label{sec:practices}
\subsubsection{The Most Commonly Installed Apps}\label{sec:most_used_apps}
\begin{figure}
\centering
\includegraphics[width=\columnwidth]{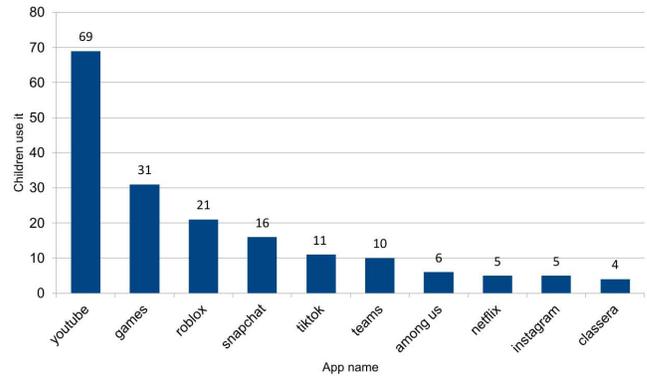}
\caption{The top 10 most used apps by children. Note that \quotes{games} is a category, not an app name, but we list it as reported by respondents.}
\label{fig:apps_rank}
\end{figure}

We asked parents to write down the two most used apps by their children.~\autoref{fig:apps_rank} shows the top 10 most used apps by children. To summarise the results, we rewrote the apps' names to unify the language (to English)\footnote{If the app name is reported in Arabic letters.}, spelling, letter case (all names are converted to small letters), and removed extra spaces. We corrected some obvious misspellings such as \quotes{Zome} which obviously refers to \quotes{Zoom}, \quotes{TikTalk} which obviously refers to \quotes{TikTok}. We removed some of the reported non-obvious apps which we could not find either in the Apple's App store or the Google Play store. It should be noted that some respondents listed app categories rather than app names such as \quotes{Games} and \quotes{Schools}. In these cases, we listed the category as reported by parents. Furthermore, we assumed that parents are aware that \quotes{YouTube} and \quotes{YouTube Kids} are two different apps (some parents reported \quotes{YouTube Kids}). We then manually checked the additional information of these apps such as the recommended minimum age, whether they require \quotes{parental guidance}, or contain violence, or request access to sensitive data\footnote{We define sensitive personal data as either contact details, location, \quotes{Device ID \& call information}, microphone, or camera data.}, from the apps' \quotes{Content rating} and \quotes{Permission} in the apps' \quotes{Additional Information} section in the Google Play Store~\cite{gplay20}. We used the Google Play store as a reference point to check the apps' additional information, regardless of the actual device brand that the child might have (e.g. iPad or Android tablet).  \par

We find that the top 10 most used apps by Saudi children are mainly Western apps, except TikTok which is Chinese. Having said that, there are 9 parents who reported local apps among the most used apps by their children. However, none of them are in our top 10 most used apps list. We checked the reported local apps in the Google Play store in December 2021. All of them fall in the education category, with one exception of an app in the entertainment category. In the education category, parents listed apps such as \setcode{utf8} \quotes{\<مدرستي>} pronounced \quotes{Madrasati}, a learning platform for school students and parents, which is owned by the Saudi Ministry of Education, \setcode{utf8}\quotes{\<دارس>} pronounced \quotes{Daress}, a platform that enables students to find private tutors, and \setcode{utf8}\quotes{\<المصحف المدرسي>} pronounced \quotes{Almus'haf Almadrasi}, an app that aims to help school students in studying and reciting Quran (the Islamic sacred book). In the entertainment category, the parent listed \setcode{utf8}\quotes{\<هادف>} pronounced \quotes{Hadif}, a local children's TV streaming app. The reasons for the limited use of local apps by children, and the effect of Western apps domination, are worth investigating in future work. \par

Comparing Saudi children's most used apps with their Western and Chinese counterparts in~\cite{zhao18,wang19}, respectively, Saudi and Western children have a similar trend, where \quotes{YouTube}, \quotes{Minecraft}, and \quotes{Roblox} (the last two are Games) are the top 3 most used apps by Western children. On the other hand, Chinese children show a different trend, where social networking apps \quotes{WeChat} and \quotes{TikTok} are the top 2 most used apps, and the third is an educational app called \quotes{NamiBox}. Unlike Saudi children, the reported top 10 most used apps by Chinese children are local: \quotes{WeChat}, \quotes{TikTok}, \quotes{NamiBox}, \quotes{Iqiyi Video}, \quotes{QQ}, \quotes{KnowBox}, \quotes{Baidu}, \quotes{Tecent Video}, \quotes{PUBG Mobile}, and \quotes{Youku Video}~\cite{wang19}. It is worth noting that China has one of the most comprehensive Internet filtering (censorship) systems (a.k.a. the Great Firewall of China~\cite{xu11}), which seems reflected on the Chinese children's most used apps ecosystem, which is dominated by local apps. Saudi Arabia is one of the countries that apply Internet filtering too~\cite{citc21}, but for specific content categories such as porn, gambling, and others~\cite{niaki21}. All of the top 10 most used apps by Saudi children are Western apps, and three of them (YouTube, Roblox, Netflix) are also listed in the most used apps by Western children as reported in~\cite{zhao18}.



\begin{table*}[!t]
\centering
\caption{The top 10 most used apps by children, and their content rating according to thr Google Play store~\cite{gplay20} except in the App named \quotes{Games} we cannot specify content rating as \quotes{Games} is a category, not an app name, and we list names as reported.}
\label{tab:apps_rank}
\resizebox{0.85\textwidth}{!}{
\begin{tabular}{lllccccc}
\toprule	
\multirow{2}{*}{App name} 	& \multirow{2}{*}{Content Rating}  & \multirow{2}{*}{\begin{tabular}[x]{@{}l@{}} Request permissions \\ to sensitive data \end{tabular}} & \multicolumn{5}{c}{Permissions} \\									
\cline{4-8}

 & & & Contact & Loc. & \begin{tabular}[x]{@{}l@{}} Device ID \& \\ Call Information\end{tabular} & Microphone & Camera \\	
											
\hline 
\multirow{2}{*}{YouTube} 	& Rated for 12+  & \multirow{2}{*}{Yes} & \multirow{2}{*}{\cmark}& \multirow{2}{*}{\cmark} & \multirow{2}{*}{\cmark} & \multirow{2}{*}{\cmark} &\multirow{2}{*}{\cmark}  \\
						    & \textbf{Parental Guidance} & & & & & & \\

\hline 
Games						& Depends on the game & N.A. &  N.A. &  N.A. & N.A. & N.A. &  N.A. \\
\hline 
\multirow{3}{*}{Roblox}		& Rated for 7+ & \multirow{3}{*}{Yes} & \multirow{3}{*}{\xmark} &\multirow{3}{*}{\xmark} &\multirow{3}{*}{\xmark} &\multirow{3}{*}{\cmark} & \multirow{3}{*}{\cmark} \\
							& \textbf{Mild Violence} & & & & & & \\
							& \textbf{Fear} & & & & & & \\
\hline 
\multirow{2}{*}{SnapChat} 	& Rated for 12+ & \multirow{2}{*}{Yes} & \multirow{2}{*}{\cmark} & \multirow{2}{*}{\cmark} & \multirow{2}{*}{\cmark} & \multirow{2}{*}{\cmark} &\multirow{2}{*}{\cmark}  \\
							& \textbf{Parental Guidance} & & & & & &  \\
\hline 
\multirow{2}{*}{TikTok}		& Rated for 12+ & \multirow{2}{*}{Yes} & \multirow{2}{*}{\cmark} &\multirow{2}{*}{\xmark} &\multirow{2}{*}{\xmark} &\multirow{2}{*}{\cmark} & \multirow{2}{*}{\cmark} \\
							& \textbf{Parental Guidance} & & & & & & \\
\hline 
MS Teams					& Rated for 3+ & Yes &  \cmark &  \cmark &  \xmark & \cmark  & \cmark \\
\hline 
\multirow{2}{*}{Among Us}	& Rated for 7+ & \multirow{2}{*}{No} &  \multirow{2}{*}{\xmark}& \multirow{2}{*}{\xmark}& \multirow{2}{*}{\xmark}& \multirow{2}{*}{\xmark}&  \multirow{2}{*}{\xmark}\\
							& \textbf{Mild Violence} & & & & & & \\
\hline 
\multirow{2}{*}{Netflix}	& Rated for 12+ & \multirow{2}{*}{Yes} & \multirow{2}{*}{\cmark}& \multirow{2}{*}{\xmark}& \multirow{2}{*}{\cmark}& \multirow{2}{*}{\cmark}& \multirow{2}{*}{\xmark} \\
							& \textbf{Parental Guidance} & & & & & & \\
\hline 
Instagram					& Rated for 12+ & \multirow{2}{*}{Yes} &\multirow{2}{*}{\cmark} & \multirow{2}{*}{\cmark}& \multirow{2}{*}{\cmark}& \multirow{2}{*}{\cmark}& \multirow{2}{*}{\cmark} \\	
							& \textbf{Parental Guidance} & & & & & & \\
\hline 
Classera					& Rated for 3+ & Yes & \xmark & \xmark & \xmark & \cmark & \cmark \\		
	
\bottomrule
\end{tabular}
} 
\end{table*}


\subsubsection{Apps' Age Appropriateness and Permissions}
\begin{figure}
\centering
\includegraphics[width=\columnwidth]{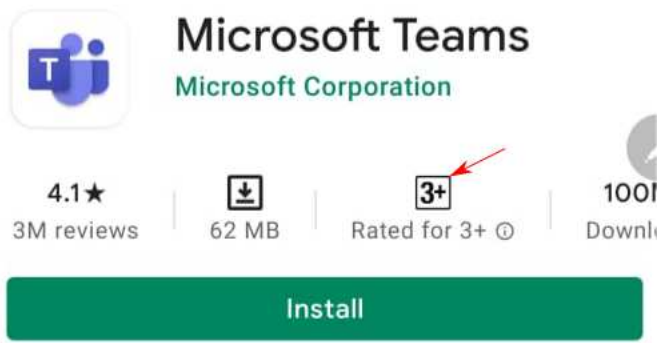}
\caption{Microsoft's \quotes{MS Teams} app appears as suitable for 3+ years old in the Google Play of an android smart phone.}
\label{fig:teams}
\end{figure}

To identify apps' age appropriateness to children aged between 6 to 12 years, we manually checked the reported top 10 most used apps against the \quotes{Content rating} section of the apps' \quotes{Additional Information} section in the Google Play store~\cite{gplay20}, using a desktop web browser. Again, we used the Google Play store as a reference point to check the apps' additional information, regardless of the actual device brand that the child might have (e.g. iPad or Android tablet). We last checked those apps on November 2021. We find 5 out of the top 10 most used apps in children's devices require \quotes{parental guidance}, while 2 others contain \quotes{mild violence}, and one contains \quotes{fear}. The first most installed app is \quotes{YouTube}, which is reported by 57.98\% of the parents. YouTube, and another four of the most used apps are appropriate only for 12+ years old children. Our results have only 23 (19.33\%) children who are 12 years old. Out of those, only 6 children who are 12 years old (5.04\%) reported YouTube as one of the most two used apps. This means that the rest of the reported YouTube users (52.94\%) are under 12 years old. Clearly, a large percentage of children are using YouTube inappropriately for their age. The second most installed app in children's devices is \quotes{Games}, which is reported by 26.05\% of the parents. The term \quotes{Games} is an app category and not a particular app. However, there are many games that may be inappropriate for children between 6 to 12 years old (our study sample), e.g. contain inappropriate content, or violence. The third most used app by children is a game called \quotes{Roblox}. While Roblox is listed in the Google Play store as suitable for children aged 7+ years old, the store states that it contains \quotes{mild violence} and \quotes{fear}. Moreover, according to the Google Play store content ratings~\cite{google21}, the content ratings are \quotes{the responsibility of the app developers and the International Age Rating Coalition (IARC)\footnote{https://www.globalratings.com}}. However, some apps' content ratings do not sound proportional to the apps' nature. For example, at the time of writing, Microsoft's \quotes{MS Teams}, an app that allows users to have video and text conversations, is listed as suitable for 3+ years old in the Google Play store, without a note for the need for parental guidance (see~\autoref{fig:teams}). \par

Second, to check the apps' permissions, we manually checked the top 10 most used apps by children against the \quotes{Permission} section of each app's details in the Google Play store~\cite{gplay20}. We last checked them on November 2021. We defined sensitive personal data as either contact details, location, \quotes{device ID \& call information}, microphone, or camera data. We then listed the apps that require permission to one or more of these data in~\autoref{tab:apps_rank}. We find that 8 out of the top 10 most used apps by children request permission to sensitive data. Apps that require access to such sensitive data need attention from parents so they give apps informed consent to have access to such data. This is especially important for parents living in countries that do not have explicit privacy laws and regulations that govern what, how, by whom, with whom, and for how long children's data can be collected, shared, and stored by apps available for children. Until September 2021, Saudi Arabia did not have personal data protection laws or regulations. However, during the peer review process of this paper, the Saudi government has issued the Personal Data Protection Law (PDPL)~\cite{boe21,alajlan21,gazette21}. According to the published official documents~\cite{boe21}, organisations were given one year (with further extensions of time to some organisations, if deemed necessary) to adjust their systems to comply with the new law~\cite{boe21}. However, recently on March 19, 2022, the Saudi Data \& Artificial Intelligence Authority (SDAIA) (who is currently in charge of implementing PDPL) issued a press release on their official account on Twitter\footnote{http://twitter.com}~\cite{sdaia_22} stating that the draft of implementing the PDPL is currently under public consultation~\cite{sdaia_22}. Subsequently on March 22, 2022, SDAIA issued a press release on their Twitter account and on the Saudi Press Agency (SPA) announcing their decision to postpone the full enforcement of the law until March 17, 2023, in order to address the comments received from stakeholders~\cite{sdaia22_2,spa22}. Therefore, at the time of writing this paper (March 24, 2022) since the PDPL has not been fully enforced yet, the announcement of the PDPL does not seem to have affected the results of this paper. However, when the PDPL is fully enforced in Saudi Arabia, we anticipate that Saudi parents' level of privacy awareness and concerns about their children's smart device apps will increase. It is worth noting that in~\cite{zhao18}, Western parents' (mostly from the UK, which already has privacy regulation) showed higher levels of privacy concerns that the Saudi parents in our study. For example, as illustrated in~\autoref{fig:installation_concerns} regarding the information parents consider when installing a new app for their children, 65\% of Western parents reported that they consider whether the app accesses any sensitive personal information, compared to 46.22\% Saudi parents. In addition, 52\% Western parents consider whether the app requires permission, compared to 40.32\% Saudi parents, and 58\% Western parents consider whether the app collects any information about their children, compared to 38.66\% Saudi parents who do so. \par

\begin{figure}
\centering
\includegraphics[width=\columnwidth]{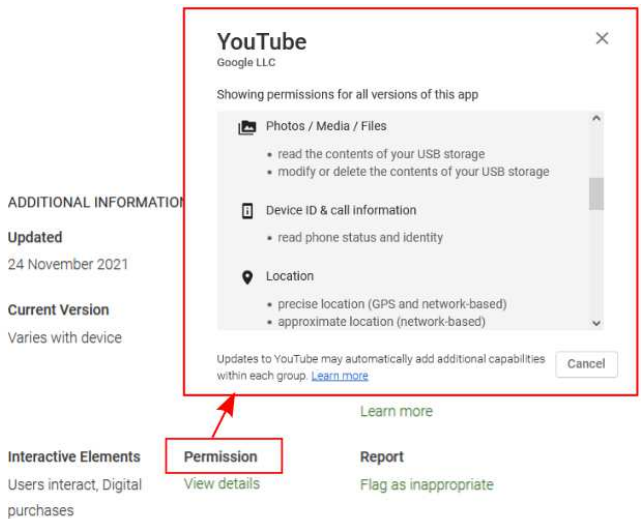}
\caption{The permissions window for an app in the Google Play store, listing all permissions that the app requests under the app's \quotes{Additional Information}.}
\label{fig:permissions}
\end{figure}

It is worth noting that most app stores such as the Google Play store communicate the apps' age appropriateness and permissions in a concise and easy to understand manner under the apps' \quotes{Additional Information} section (or similar sections in other stores), which is outside the usually long and complicated terms and conditions page. The additional information about an app can be viewed easily before installing an app. For example,~\autoref{fig:permissions} shows the permissions an app requests access to. Having said that, some users may be unaware of the availability of such information. This is one of the issues that needs education efforts from security and privacy experts to ordinary users. Thus, we listed the need for effective education in the recommendations section~(see~\autoref{sec:recomm}).

\subsubsection{Apps Installation} \label{sec:installation}
We asked parents how their children installed their most used apps. We listed multiple choices and asked parents to select all that apply (i.e. parents could select more than one choice). 57.14\% of the parents said that apps were installed by the parent(s) after the child asked for them, 32.77\% said that apps were installed by the child himself/herself, while only 11.76\% said that they installed the apps after some research.~\autoref{tab:installation} shows the detailed results. \par 

\begin{table*}[!t]
	\centering
	\caption{How children's most used apps are installed on their devices. The last two choices were not included in the Western and Chinese parents' studies, hence their columns are \quotes{-}.}
	\label{tab:installation}
	\begin{tabular}{ll@{\hspace{5.5pt}}rrr}
		\toprule
		\multirow{2}{*}{Installation method} & \multicolumn{4}{c}{Parents} \\
		\cline{2-5}
		& \multicolumn{2}{c}{Saudi} & Western & Chinese \\ 
		\hline
		\rowcolor{lightYellow}
		My child asked for them, and then I installed them for him/her 
		& 68 &(57.14\%) & 71\% & 30.7\% \\
		I found the app after some research, and installed it for him/her 
		& 14 &(11.76\%) &  15\% & 20.1\%\\
		I found the app following some recommendations, and installed it for him/her 
		& 17 &(14.29\%) & 8\% & 7.9\% \\
		\rowcolor{lightYellow}
		My child might have installed them by him/herself  
		& 39 & (32.77\%) & 19\% & 28.6\% \\
		I do not know 
		& 0 & (0\%) & - & - \\
		Other 
		& 8 &(6.72\%) & - & - \\
		\bottomrule
	\end{tabular}
\end{table*}


Comparing how Saudi children's favorite apps are installed with their Western and Chinese counterparts in~\cite{zhao18,wang19}\footnote{It should be noted that our study and the one on Western parents~\cite{zhao18} allowed participants to select all that apply, while the study on Chinese parents~\cite{wang19} appears to allow a single choice as the sum of all answers is exactly 100\%.} respectively, we find interesting patterns. First, 32.77\% of Saudi parents reported that their children's apps were installed by the children themselves, compared to only 19\% of Western parents, and 28.6\% of Chinese parents who said so. At first glance, this might reflect a higher level of autonomy by both Saudi and Chinese children than their Western counterparts. However, such autonomy in apps installation at this age can also be due to parents' low technical background, or due to low level of privacy awareness by Saudi and Chinese parents. That is, when parents are not aware of the privacy risks associated with smart devices' apps, they may give their children more autonomy to install apps. The results suggest that Western parents are more concerned about privacy issues than Saudi and Chinese parents, which makes most Western parents install their children apps, possibly to ensure the apps' safety for their children first. Second, Saudi parents appear to be more influenced by recommendations concerning children's apps selection (14.29\%), compared to 8\% of Western parents and 7.9\% of Chinese parents. Third, Saudi parents are the least likely to select their children apps after some research (11.76\%), compared to 15\% of Western and 20.1\% of Chinese parents. 

\subsubsection{Apps Installation Concerns}
\begin{figure*}
\includegraphics[width=0.9\textwidth]{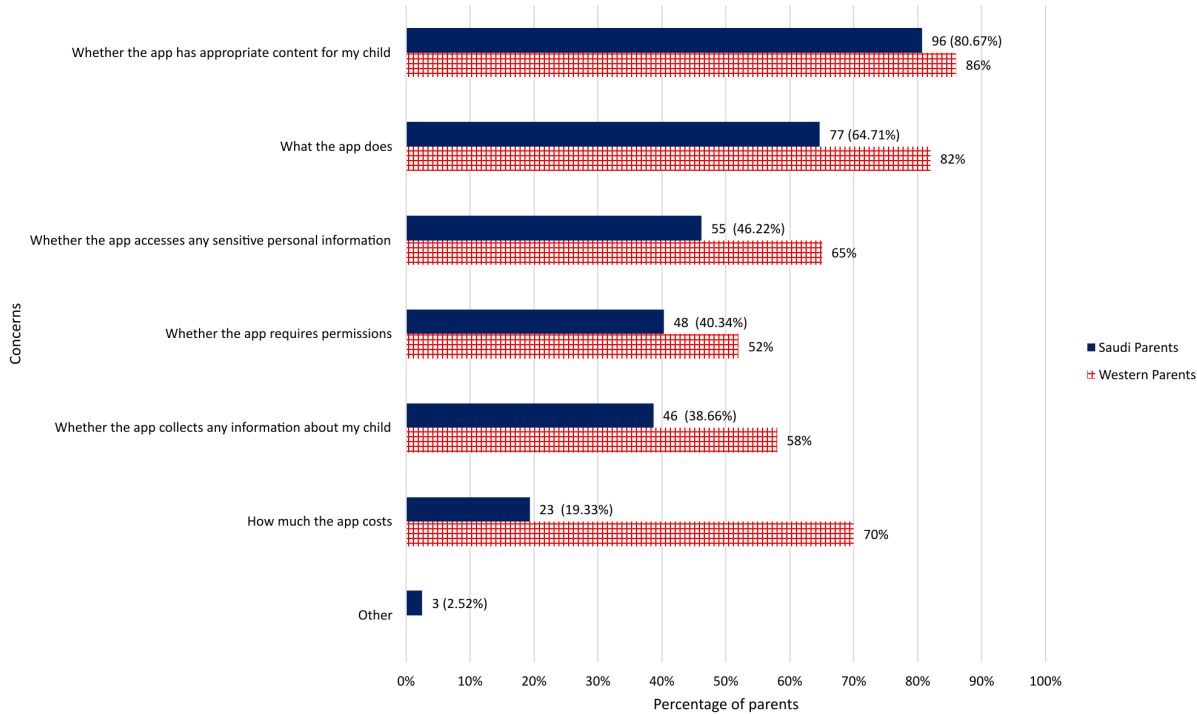}
\caption{Information considered by the Saudi vs. Western parents when installing apps for their children.}
\label{fig:installation_concerns}
\end{figure*}

With respect to parents' concerns when installing their children's apps, we asked parents about the information they consider when installing a new app for their children. We allowed parents to select all that apply. As illustrated in~\autoref{fig:installation_concerns}, the app's content is the highest concern among Saudi parents, and is selected by 80.67\%, followed by what the app does, which is selected by 64.71\%. Parents' concern about an app's access to sensitive personal information appears as the third most selected concern, where it is selected by less than half of the parents (46.22\%), followed by whether the app requires permissions (40.34\%) and whether the app collects any information about their child (38.66\%). These figures indicate that parents are more concerned about what their children receive (app's content and function) than what their children provide to the app (app's access to sensitive personal information, permissions, or the information the app collects about the child). Therefore, awareness for parents and their children about the importance of both aspects is needed. Finally, our results also show that the app's cost comes as the last concern by Saudi parents, selected by only 19.33\% parents. The lack of concern about apps' costs by Saudi parents could be related to the lack of financial skills that is often observed in the Saudi society~\cite{alghamdi21,almushare15,kpmg20}.\par 

Comparing Saudi parents' apps installation concerns with those expressed by Western and Chinese parents in~\cite{zhao18,wang19}, as illustrated in~\autoref{fig:installation_concerns}, overall, Western parents show higher levels of concern than Saudi parents, in every aspect. 70\% Western parents were concerned about the app's cost compared to 19.33\% Saudi parents. Similarly, 58\% Western parents are concerned about whether the app collects any information about their child compared to 38.66\% Saudi parents, and 65\% Western parents are concerned about whether the app accesses any sensitive personal information compared to 46.22\% Saudi parents. Roughly speaking, Chinese parents show a similar trend to Saudi and Western parents\footnote{The comparison of Saudi parents' installation concerns for their children's apps with Chinese parents in~\cite{wang19} is a high-level comparison. This is because the answer choices provided to the respondents in~\cite{wang19} are not identical to ours. However, both have similarities that allow us find some high-level patterns.}. That is, privacy concerns (such as sharing and disclosing personal information) are not the highest concerns, and inappropriate content concern comes higher than privacy. \par
Having said that, even though Western parents in~\cite{zhao18} selected more concerns than Saudi parents, both Western and Saudi parents share the highest two concerns: the app's content, and what the app does. This suggests that parents' concern about the information their children receive more than the information that their children provide or the information that the app collects is a global issue.


\subsubsection{Password Protection}
We asked parents if they have a password to log in to the device or to install apps on their children's devices, for example, to require their children to ask them before they can log in to the device or install apps. Our results show 84 (70.59\%) of parents have a password on their children's devices, while the remaining 35 (29.41\%) do not have any. The percentage of parents who do not have passwords protection in their children's devices is somewhat high: around a third of the parents do not use passwords in their children's devices neither to log in to the device nor to install apps. When we asked parents if they have any other mechanisms in place to protect their children's privacy during their interaction with the device only 40 (33.61\%) answered \quotes{yes}. Moreover, of the 35 parents who do not have a password, 28 (80\%) reported that they do not have any other mechanism. For those who do not use a password, without any alternative access control mechanism, the absence of login passwords represents a threat to children's privacy, especially in case of lost or stolen devices. Login passwords provide access control and protect children's  privacy especially in case of lost or stolen devices, while app installation passwords provide control over free and paid apps installation. However, there is a trade-off between privacy and usability. Using passwords provides privacy benefits to control access and apps installation, but sometimes, at the cost of usability. For example, in the app installation passwords that are set and owned by parents, children may try to figure out the password, or even worse, change the password, to overcome their parents' control. Moreover, sharing passwords might conflict with parents' privacy.  \par

\begin{figure*}
	\centering
	\includegraphics[width=0.9\textwidth]{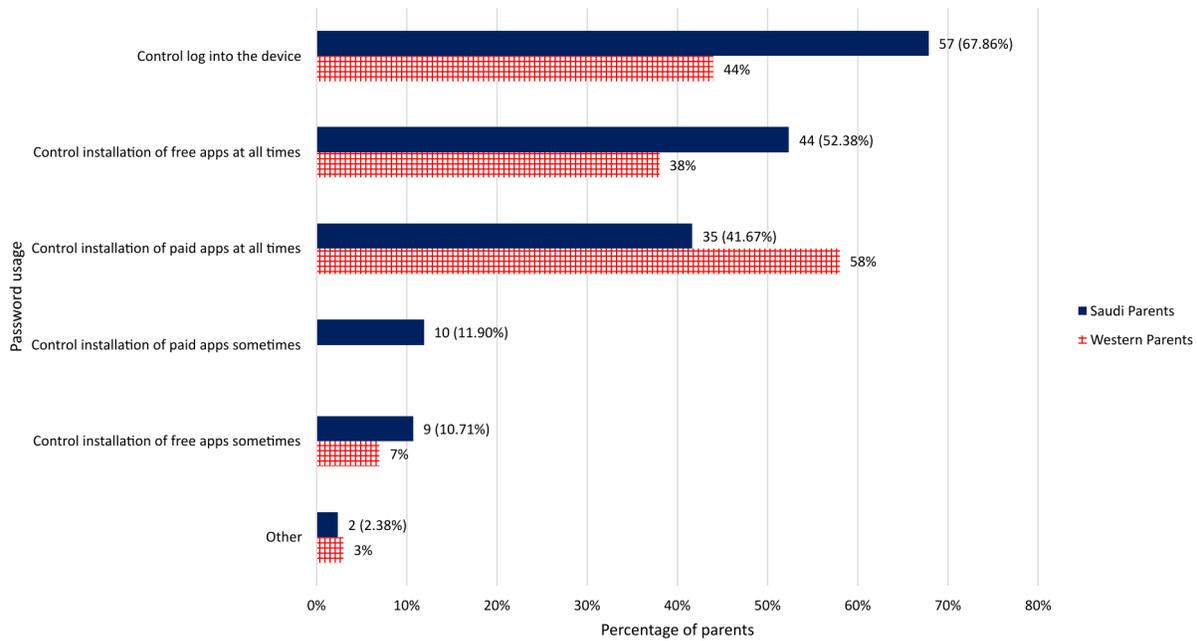}
	\caption{Reasons for having a password on children's devices by Saudi vs. Western parents. Note that the option \quotes{Control installation of paid apps sometimes} does not exist in the Western survey, hence the percentage is not present in this option for the Western parents.}
	\label{fig:password_reasons}
\end{figure*}

Out of those who use passwords, we asked them how they use passwords in their children's devices, and we allowed them to select all that apply. 67.86\% of the parents use it to control who can log in to the device, and 52.38\% use it to control installation of free apps at all times, and 41.67\% use it to control installation of paid apps at all times. See~\autoref{fig:password_reasons} for the full results. These figures show that the first motivation to have a password in the child's device is to control login to the device, then to control installation of free, then paid apps at all times. While 67.86\% of the parents who use passwords use it to control access (login) to their children's devices, this percentage shows a lack of using passwords to control access to children's devices by more than a third (32.14\%) of the parents who use passwords. \par

Compared to Western parents in~\cite{zhao18}, there is a considerably high percentage of Western children's devices without passwords too (24\%). However, Western parents who use passwords have different priorities than Saudi parents. As illustrated in~\autoref{fig:password_reasons}, the first motivation for Western parents to have a password is to control the installation of paid apps (58\%), followed by controlling login to the device (44\%), then controlling the installation of free apps at all times (38\%). This may suggest that Western parents' main motivation is influenced by financial reasons (controlling paid apps). On the other hand, Saudi parents' main motivation is for access control and controlling free apps, while control installation of paid apps comes as the third most selected reason.

\subsubsection{Refusing to Install, or Uninstalling, an App}
When we asked parents if they have ever refused to install an app for their children because of privacy concerns about the app, 102 (85.71\%) of parents answered \quotes{yes}. Moreover, we asked parents if they have ever uninstalled an app for their children because of privacy concerns, 98 (82.35\%) of parents answered \quotes{yes}. Overall, 109 (91.60\%) of Saudi parents have refused to install, or uninstalled, an app from their children's devices, because of privacy concerns about an app. \par 

Comparing Saudi parents' answers with their Western counterparts in~\cite{zhao18}, we find that Saudi parents show a different trend from Western parents, where only 67\% of Western parents said that they had refused to install an app, and only 54\% had uninstalled an app for their children because of privacy concerns. However, looking at the previously mentioned figures in apps installation concerns in~\autoref{tab:installation}, we observe higher percentages of Saudi children (32.77\%) who install their apps by themselves, compared to only 19\% of Western children who do so. Moreover, we observe lower percentages of Saudi parents (57.14\%) who install apps for their children after their children ask for them, compared to 71\% of Western parents who do so. This suggests that parents' involvement in installing their children's apps is related to a reduced number of children's apps uninstallation incidents.\par

Wang et al.~\cite{wang19} asked a similar question\footnote{It should be noted that it is not clear from Wang et al.'s report~\cite{wang19} if their questions about whether parents had refused to install, or uninstalled, an app for their children included the sentence \quotes{because you had any privacy concerns about the app} or was open to any reason. However, they stated earlier in their report that \quotes{questions in the survey were based on our previous online surveys with UK parents~\cite{zhao18}, and related literature}. The UK survey~\cite{zhao18} questions about whether parents had refused to install, or uninstalled, an app for their children contained the sentence \quotes{because you had any privacy concerns about the app}, which we included in our survey questions too.} and found that 98\% of Chinese parents had refused to install, or uninstalled, an app for their children. Wang et al.~\cite{wang19} interpreted such differences between the Chinese and Western parents' practices by the \quotes{authoritarian parenting style}, which is often observed among Chinese parents~\cite{wang19}. Such interpretation can be applicable, to some extent, to Saudi parents too as an Eastern society. In~\cite{baumrind66} Baumrind defined three parenting styles: the \quotes{authoritarian}, the \quotes{permissive}, and the \quotes{authoritative}. In the \quotes{authoritarian} parenting style, parents exercise control to enforce a set of standards, which are usually motivated by a higher authority, and expect the children's unquestioned adherence to them. Contrary to this, is the \quotes{permissive} parenting style, in which parents avoid exercising control. They represent themselves as a source for the children that the children can refer to when they need. They enable the children to regulate their own activities. The \quotes{authoritative} parenting style represents a middle ground between the two extremes (the authoritarian and the permissive parenting styles), in which parents exercise rational control, discuss the reasoning behind their decisions, and enable gradual autonomy to the children. However, according to Dwairy et al.~\cite{dwairy06} \quotes{the parenting styles among Arabs are not as distinct as in the West}. Thus, Dwairy et al. defined several parenting \quotes{patterns}, and found that the \quotes{controlling pattern}, which is defined as the combination of the \quotes{authoritarian} and \quotes{authoritative} parenting styles, is prevalent in Saudi Arabia. The \quotes{authoritative} parenting style is known to be prevalent in Western societies~\cite{dwairy06}.\par

\subsubsection{Reasons for Refusing to Install, or Uninstalling, an App}
Of those parents who have refused to install, or uninstalled, an app for their children, we asked them to select the concerns that led them to do so. We listed several concerns and asked them to select all that apply. We find that 91 (83.49\%) parents selected \quotes{inappropriate content}, while only 49 (44.95\%) selected \quotes{the app was asking for access to too many things such as camera, location, or microphone...}, and 39 (35.78\%) selected \quotes{the app was accessing too much personal information about my child}. \par 

\begin{figure*}
	\includegraphics[width=0.9\textwidth]{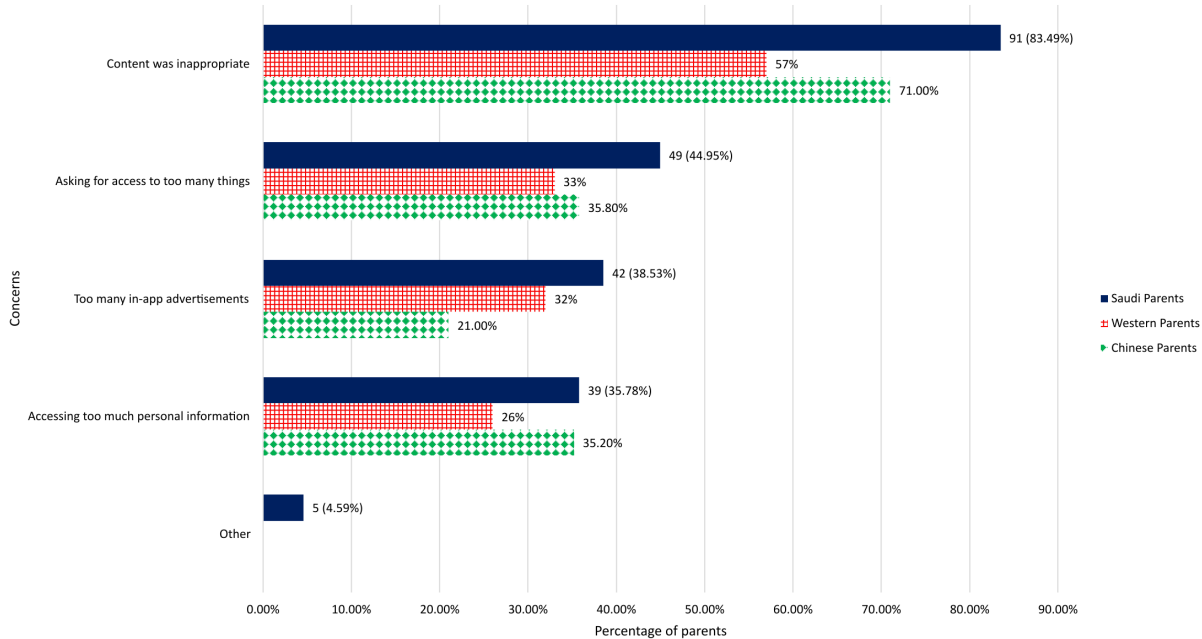}
	\caption{Saudi vs. the Western vs. Chinese parents' concerns that caused refusing to install, or uninstall, an app for their children.}
	\label{fig:refuse_uninstall_concern}
\end{figure*} 
Comparing Saudi parents' reasons for refusing to install, or uninstalling, an app for their children with their Western and Chinese counterparts in~\cite{zhao18,wang19} respectively, as~\autoref{fig:refuse_uninstall_concern} illustrates, we find that they all show a similar trend. That is, in all the three societies, the primary concern for refusing to install, or uninstalling, an app is \quotes{inappropriate content} (information the child receives), while they show less concern about privacy (information the child provides or that the app accesses). However, the figures show lower concern about inappropriate content by Western parents, where only 57\% have concerns about it, compared to 83.49\% of Saudi parents, and 71\% of Chinese parents. This suggests that classifying content as \quotes{inappropriate} may vary between cultures. Such variations worth considering by parents, developers, and regulators with respect to children's online safety tools and content rating schemes. \par

In this regard, we recently became aware of an initiative in Saudi Arabia called \quotes{Qayyem}~\cite{qayyem21}, which is a step towards bridging the cultural gaps in the general rating schemes. However, it is for online games (not mobile app) which are mostly operated on Play Station, XBOX, and Personal Computer (PC) devices. Qayyem initiative is initiated by Prince Muhammed bin Salman Foundation (Misk)~\cite{misk21}, a non-profit organization which aims to \quotes{cultivate and encourage learning and leadership in youth for a better future in Saudi Arabia}~\cite{misk21,arabnews21}. Qayyem provides a rating scheme for children's most popular online games. It is not limited to Arabic games. In fact, all the games we could see on their website are in English~\cite{qayyem21}. The games are rated by expert players who undergo training about the rating criteria for Qayyem's scheme~\cite{qayyem_faq21}. The players' notes are also reviewed by a specialised committee~\cite{qayyem_faq21}. The scheme considers evaluating content and age rating. The content is evaluated against five criteria~\cite{qayyem21}: \begin{inparaenum}\item \quotes{Religious Remarks}, to report concerns if the game has content that conflicts with Islamic values. \item \quotes{Psychological Remarks}, to report concerns if the game might affect the player's psychological aspect negatively. \item \quotes{Behavioural Remarks}, to report concerns if the game might affect the player's \quotes{clear-headedness}. \item \quotes{Moral Remarks}, to report concerns if the game has content that conflicts with \quotes{decency, morality and good behaviour}. \item \quotes{Financial Remarks}, to report concerns if the game conflicts with financial and monetary values and integrity\end{inparaenum}. For example, we looked at one rated game called \quotes{Marvel's Avengers}, and we find several remarks~\cite{marvels21}. First, a religious remark with \quotes{low} severity, raised a concern regarding a reference to \quotes{Thor}, which is the God of thunder in the Norse mythology. Second, a psychological remark with \quotes{medium} severity, raised a concern regarding repeated virtual violence where \quotes{the player fights soldiers and robots, the defeated falls, then disappears, without any blood appearing}. Third, a behavioural remark with \quotes{medium} severity, raised a concern regarding an alcohol drinking scene. Fourth, a  moral remark with \quotes{low} severity, raised a concern regarding some actors appear with too tight clothes, and the use of bad words. The games' ratings appear only in the Arabic version of the website, although all the rated games we can see are in English. Our observation is that most of Qayyem's content rating criteria are normally considered in children's local media content and public school educational materials, and are generally widely respected by the Saudi society. Having said that, we do not have data to evaluate this initiative's effectiveness or its penetration rate among Saudi families. Moreover, it is not clear what Saudi families' reactions would be regarding Qayyem's remarks, if their children installed or want to install a game with remarks. Will parents refuse to install, or uninstall, a game because of Qayyem's rating remarks? or will they just raise their children's awareness about cultural differences and let them play it despite the remarks? The initiative provides these ratings for parents, but leave the choice for them as they state: \quotes{we played our part and rated it for you[,] play your part and pick the right one for them}~\cite{qayyem21}. Moreover, Qayyem has a disclaimer that states that their ratings are considered \quotes{personal, independent, and non-binding opinions}~\cite{qayyem21}.

\subsubsection{Children's Attitudes Towards Being Refused to Install an App, or Having an App Being Uninstalled by Parents}
We asked parents about how their children felt about being refused to install an app or having their app uninstalled. To summarise the responses we manually labelled the answers with either: \quotes{positive}, \quotes{neutral}, or \quotes{negative}, based on the reported attitude. For example, answers such as \quotes{She was very understanding after I explained to her the risks} were labelled \quotes{positive}, answers that contain words such as \quotes{sad}, \quotes{mad}, and \quotes{upset} were labelled \quotes{negative}. Answers that contain negative then positive attitudes such as \quotes{she was upset at the beginning, and after explaining the reasons, she accepted the issue} were labelled negative too. Answers such as \quotes{lots of questions and demand to explain the reasons} were labelled \quotes{neutral}. This method provides us with an estimation. \par 

Out of the 109 parents who refused to install, or uninstalled, an app for their children, we find that $\sim$63 children's reactions are negative (57.80\%), while around $\sim$31 (28.44\%) are positive, and around $\sim$15 (13.76\%) are neutral. To reduce the conflicting situation of uninstalling apps for children, which was mostly negatively received by children, we suggest parents' early involvement during the apps selection and installation to decide on the app appropriateness before the child has it. However, experience shows that parents' involvement and conversations with children using commonly-understood language might be non-trivial due to several reasons, such as lack of suitable educational resources for the parents themselves. \par

Comparing Saudi parents with their Western and Chinese counterparts in~\cite{zhao18,wang19} respectively, similar to Saudi children, according to~\cite{zhao18}, out of those Western parents who had uninstalled an app for their children, a total of 55\% reported that their children reacted negatively (\quotes{upset and could not understand their parents' decision} (48\%), \quotes{sad} or \quotes{confused} (7\%)) when parents uninstalled an app for them, while still 40\% reported that their children were \quotes{capable of understanding parents' decisions}, and 6\% reported that their children \quotes{were not bothered}. In comparison to Chinese parents, Wang et al.~\cite{wang19} asked Chinese parents a different but somewhat relevant question. They asked parents about their children's reactions to \quotes{parents safeguarding practices}. 58.3\% of Chinese parents said their children \quotes{would understand parents' decisions after explaining the risks to them}, 12.1\% said that their children \quotes{would fully understand the decisions and accept without any argument}, 22.1\% of parents said that their children \quotes{would be able to understand them but refused due to some other reasons}, while only 7.1\% parents said that their children \quotes{would be upset and refuse their parents' help}. Wang et al.~\cite{wang19} linked this to the \quotes{authoritarian parenting style}, which is often observed in Chinese parents.

\subsection{Privacy Concerns}
\subsubsection{Thinking and Talking about Privacy with Children}\label{sec:think_talk_priv}
We asked parents how often they think about their children's online privacy in relation to their interaction with tablet or smartphone devices. Saudi parents seem very concerned about their children's online privacy. 52.94\% of parents reported that they think about it \quotes{at least once a day}, and 27.73\% of parents think about it \quotes{at least once a week}. See~\autoref{tab:think_talk_privacy} for the rest of the answers. \par 

We then asked parents how often they talk about online privacy issues with their children in relation to interactions with tablet or smartphone devices. We found that 36.97\% of parents reported that they talk about it \quotes{at least once a week}, 26.05\% of parents talk about it \quotes{at least once a month}, and 14.29\% of parents talk about it \quotes{at least once a day}. Having said that, there are 7.56\% who \quotes{do not talk about privacy at all} with their children, which is a negative point. However, it can be an indicator of parents' lack of awareness or confidence to talk about privacy issues with their children.~\autoref{tab:think_talk_privacy} shows how often parents think and talk with their children about online privacy issues. \par

\begin{table}[!t]
	\centering
	\begin{threeparttable}
	\caption{Frequency of parents' thinking versus talking with their children about children's online privacy}
	\label{tab:think_talk_privacy}
	\begin{tabular}{lll@{\hspace{5pt}}rl@{\hspace{5pt}}r}
		\toprule
		\multirow{2}{*}{Frequency} & & \multicolumn{2}{c}{Think} & \multicolumn{2}{c}{Talk} \\
		 \cline{3-6}
			& & \multicolumn{2}{c}{Parents} & \multicolumn{2}{c}{Parents} \\
		\hline 
		At least once a day & &   63 & (52.94\%) & 17 & (14.29\%)\\ 
		At least once a week & &  33 & (27.73\%) & 44 & (36.97\%)\\
		At least once a month & & 11 & (9.24\%) & 31 & (26.05\%) \\
		At least once every couple of months & & 3 & (2.52\%) & 14 & (11.76\%)\\
		At least once a year & & 1 & (0.84\%) & 2 & (1.68\%)\\
		Less than once a year & & 0 & (0\%) & 2 & (1.68\%)\\
		I do not (think $\vert$ talk)\mtnote{1} about it at all & & 5 & (4.20\%) & 9 & (7.56\%)\\
		Other & & 3 & (2.5\%) & 0 & (0\%) \\
		\bottomrule	
	\end{tabular}
		\begin{tablenotes}
		\item[1] Think or talk, depending on the question.
		\end{tablenotes}
	\end{threeparttable}
\end{table}
Comparing how Saudi parents think and talk about privacy with their children against their Western and Chinese counterparts in~\cite{zhao18,wang19}, we find that Western parents expressed high level of concern about their children's privacy too, where 45\% of them said they think about it \quotes{very often}, and 47\% \quotes{sometimes} or \quotes{occasionally} do so\footnote{Note that, Zhao et al.~\cite{zhao18} and Wang et al.~\cite{wang19} used different scales than ours to measure frequency. However, our comparison is at a high level which does not require exact matching.}. Contrary to both Saudi and Western parents, Chinese parents did not express a high level of concern about their children's privacy. As reported in Wang et al.~\cite{wang19}\footnote{Wang et al.~\cite{wang19} asked about parents' level of concern in a general sense and not particularly about the frequency of talking about privacy. However, this question is the closest to our comparison which is at a high level and does not require exact matching.}, only 10\% of Chinese parents have \quotes{a lot of} concerns about their children's online privacy, 47.5\% have \quotes{some but acceptable}, 32.4\% have \quotes{no} or \quotes{very little}, and 8.3\% had never thought about this issue. The latter is a striking figure compared to only 4.20\% of Saudi parents who do not think about their children's privacy at all. The low level of privacy concerns that is observed in Chinese parents could be related to cultural or political reasons.

Regarding talking about privacy issues with children, 62\% of Chinese parents \quotes{had never} or \quotes{very rarely} discussed privacy issues with their children. Again, this is a striking figure compared to Saudi parents, where the majority discuss privacy issues with their children regularly as shown in~\autoref{tab:think_talk_privacy} (second column). We could not compare Saudi parents to Western parents in this aspect (talking about privacy with their children) as the report on Western parents~\cite{zhao18} does not contain this data. 

\subsubsection{Parents' Level of Privacy Concerns Before versus After Being Informed About Possible Privacy Implications}
\begin{figure}
	\centering
	\includegraphics[width=\columnwidth]{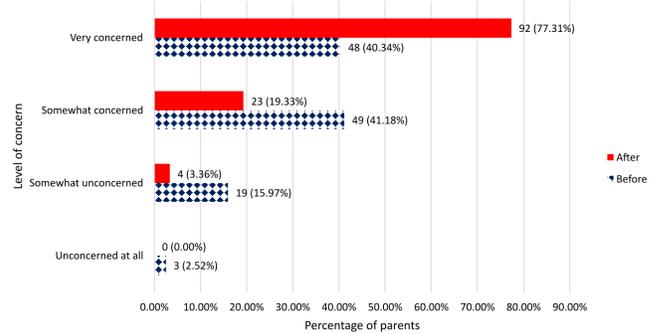}
	\caption{Parents' level of concern regarding apps' access to camera and microphone \textit{before} and \textit{after} being informed about possible privacy implications.}
	\label{fig:camera_concern}
\end{figure}

\begin{figure}
	\centering
	\includegraphics[width=\columnwidth]{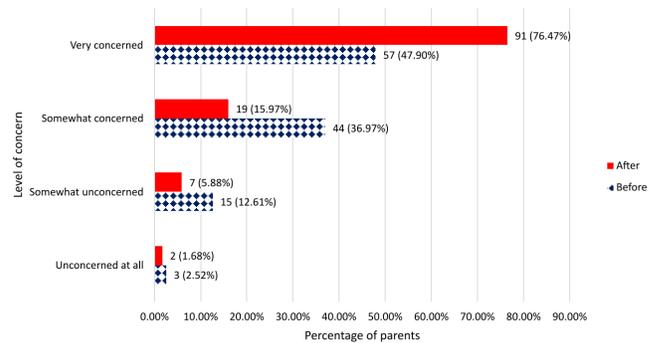}
	\caption{Parents' level of concern regarding apps' access to location \textit{before} and \textit{after} being informed about possible privacy implications.}
	\label{fig:location_concern}
\end{figure}

We asked Saudi parents about their level of concern if an app used by their children asked for access to the camera, microphone, or location service on the device. Most Saudi parents expressed high levels of concerns about apps that request access permission to their children's camera, microphone (77.31\% are \quotes{very concerned}), or location service (76.47\% are \quotes{very concerned}). However, their levels of concern significantly increased when they were informed about possible privacy implications of giving access permissions to their child's camera (or photos), microphone, or location. This was revealed in scenario-based questions which asked about the same set of permissions differently. See~\autoref{tab:concerns} for the two questions we ask (\textit{before} and \textit{after} being informed about possible privacy implications).~\autoref{fig:camera_concern} and \autoref{fig:location_concern} illustrate the results. \par 

After being informed about possible privacy implications, the percentage of the \quotes{very concerned} parents about their children's privacy with respect to camera and microphone access permission increased from 40.34\% to 77.31\%. Similarly, the percentage of the \quotes{very concerned} parents with respect to location services increased from 47.90\% to 76.47\%. See~\autoref{fig:camera_concern} and \autoref{fig:location_concern} for detailed numbers.


\begin{table*}[!t]
\centering
\caption{The questions asked to parents about their level of parents' concerns: \textit{before} and \textit{after} we state possible privacy implications in the question.} 
\label{tab:concerns}
\begin{tabularx}{\linewidth}{cLL}
\toprule
Serial No. & Question without privacy implications (\textit{before}) & Question with privacy implications (\textit{after})\\
\hline 
1 & How concerned would you be if an app used by your child asked for access to the camera or microphone on the device? 
& Many apps have access to a device's photos or microphone, which could enable strangers to ask your child to share their photos or talk with them. How concerned are you about this possibility?\\
\hline
2 & How concerned would you be if an app used by your child asked for access to the location service on the device? 
& Some apps can have access to a device's location information, which could enable other organisations/companies to infer which school your child goes to, when and where you go on holiday, etc. How concerned are you about this possibility?\\
\bottomrule
\end{tabularx}
\end{table*}

\begin{table*}
\centering
\caption{Saudi versus Western versus Chinese parents' level of concerns regarding apps' access to microphone, camera, and photos permissions before and after knowing the implications. Columns with \quotes{-} value mean we do not have figures for them.}
\label{tab:camera_concerns_ba}
\begin{threeparttable}
\begin{tabular}{lll@{\hspace{5.5pt}}rl@{\hspace{5.5pt}}rll@{\hspace{5.5pt}}rl@{\hspace{5.5pt}}rll@{\hspace{5.5pt}}rl@{\hspace{5.5pt}}r}
\toprule 
\multirow{3}{*}{Level of concern about}  & &  \multicolumn{14}{c}{Parents}\\
\cline{3-16}\\
Mic/Camera/Photo permission & & \multicolumn{4}{c}{Saudi} & & \multicolumn{4}{c}{Western} & & \multicolumn{4}{c}{Chinese}\\
	\cline{3-6} \cline{8-11} \cline{13-16}\\
 		
 & &   \multicolumn{2}{c}{Before} & \multicolumn{2}{c}{After} & & \multicolumn{2}{c}{Before} & \multicolumn{2}{c}{After} & & \multicolumn{2}{c}{Before} & \multicolumn{2}{c}{After} \\
\hline

Very concerned  	  & \vline 
					  & 48 & (40.34\%) & 92 & (77.31\%) &\vline 
					  & -  & (51\%)   & -  & (75\%)   & \vline 
					  & -  & (23.3\%) & -  & (39.9\%) \\
					  
Somewhat concerned    & \vline 
					  & 49 & (41.18\%) & 23 & (19.33\%) & \vline
				      & - & (39\%) & - & (24\%) &\vline 
				      & - & (36.9\%) & - & (33.7\%) \\
				      
Somewhat unconcerned\tnote{1}  
					  & \vline 
					  & 19 & (15.97\%) & 4 & (3.36\%) &\vline 
					  & - & (9\%) & - & (1\%) &\vline 
					  & - & (21.7\%) & - & (16\%) \\
					  
Unconcerned at all    & \vline 
					  & 3  & (2.52\%)  & 0 & (0\%) & \vline
					  & - & (1\%) & - & (0\%) &\vline 
					  & - & (11.8\%) & - & (5.6\%) \\
Don't know/never though about this 	 
					  & \vline 
					  & -  & -  & - & - & \vline
					  & - & - & - & - &\vline 
					  & - & (6.3\%) & - & (4.8\%)  \\				  
\bottomrule	
\end{tabular}
\begin{tablenotes}
\item[1] In Western parents~\cite{wang19} and Chinese parents studies~\cite{zhao18}, this metric is \quotes{Not too concerned}, which we match with our \quotes{Somewhat unconcerned}.
\end{tablenotes}
\end{threeparttable}
\end{table*}

Comparing Saudi parents' level of concern about their children's privacy before and after being informed about possible privacy implications against their Western and Chinese counterparts in~\cite{zhao18,wang19}, we find that Western parents show a similar trend to Saudi parents. As shown in~\autoref{tab:camera_concerns_ba}, 51\% of Western parents are \quotes{very concerned} about microphone and photo access, and this percentage increased to 75\% after being informed about possible privacy implications. However, regarding concerns about location access, there is not much increase in the level of concern after being informed about possible privacy implications since the majority of Western parents (62\%) are already \quotes{very concerned} about location, and this percentage increased to 69\%, after being informed about possible privacy implications. However, this is not the case with Chinese parents, where Wang et al.~\cite{wang19} reported much lower levels of concerns expressed by Chinese parents even after being informed about possible privacy implications. 23.3\% are \quotes{very concerned} about microphone and photo access, and this percentage increased to 39.9\%, after being informed about possible privacy implications. Moreover, 40.2\% who are \quotes{very concerned} about location access, and this percentage increased to 48.7\%, after being informed about possible privacy implications. Clearly, the Chinese parents' level of concern about their children's privacy is lower than that expressed by Saudi and Western parents. This could be related to cultural or political reasons as we stated earlier in~\autoref{sec:think_talk_priv}.

\subsection{Parental Controls}
\subsubsection{Parents' Experience With Privacy Settings}
We asked parents if they have heard about privacy permission settings for an app and if so, how often they check privacy permission settings for their children's devices. Our results show that 87 (73.11\%) of Saudi parents have heard about privacy permission settings for an app. Of those, a total of 45.98\% reported that they check privacy permission settings for their children's devices at least \quotes{once a month}, \quotes{at least once a week}, or \quotes{at least once a day}. See~\autoref{tab:frequency_checking} for detailed results. Of those who check privacy settings, 90.79\% either \quotes{strongly agree} or \quotes{somewhat agree} that tuning privacy permission settings reduces their privacy concerns about what their children use on the device. See~\autoref{tab:permissions} for detailed results. Interestingly, 100\% of those who have not heard about privacy permissions for an app are interested to learn more about them.\par 

Comparing Saudi parents' awareness and frequency of checking privacy permission settings for their children's apps with their Western counterparts in~\cite{zhao18}, we find that Western parents show similar results. 72\% of Western parents have heard about privacy permission settings. Moreover, a total of 66\% of Western parents check privacy permission settings for their children's devices either \quotes{from time to time} or \quotes{very regularly}, and 24\% of them check those settings \quotes{whenever a new app is installed}. However, there are 11\% who \quotes{can't remember} when was the last time they checked them, compared to only 8.05\% of Saudi parents who said so. Similar to Saudi parents, the majority of Western parents who check privacy settings (78\%) felt that this reduces their privacy concerns about what their children use on the device. We could not compare Saudi parents to Chinese parents in this aspect (awareness and frequency of checking privacy permission settings for their children's apps) as the report on Chinese parents~\cite{wang19} does not contain this data.

\begin{table}[!t]
	\centering
	\caption{Frequency for checking privacy settings.}
	\label{tab:frequency_checking}
	\begin{tabular}{ll@{\hspace{5pt}}r}
		\toprule
		Frequency & \multicolumn{2}{c}{Parents} \\
		\hline 
		At least once a day & 7 & (8.05\%) \\
		At least once a week & 19 & (21.84\%) \\ 
		At least once a month & 14 & (16.09\%)\\
		At least once every couple of months & 20 & (22.99\%)\\
		At least once a year & 0 & (0\%) \\
		Less than once a year & 4 & (4.60\%)\\
		Whenever a new app is installed & 12 & (13.79\%) \\
		I don't remember. Last time was a long time ago & 7 & (8.05\%)\\
		I don't check them at all & 4 & (4.60\%)\\
		\bottomrule
	\end{tabular}
\end{table}

\begin{table}[!t]
	\centering
	\caption{Parents' level of agreement towards Q. 46 statement: \quotes{I feel that tuning the privacy permission settings reduces my privacy concerns about what my child uses on the device}.}
	\label{tab:permissions}
	\begin{tabular}{ll@{\hspace{5pt}}r}
		\toprule
		Level of agreement & \multicolumn{2}{c}{Parents} \\
		\hline 
		Strongly agree & 30 & (39.47\%)\\
		Somewhat agree & 39 & (51.32\%)\\
		Neither agree nor disagree (neutral) & 5 & (6.58\%)\\
		Somewhat disagree & 2 & (2.63\%) \\ 	  
		Strongly disagree & 0 & (0\%)\\
		\bottomrule
	\end{tabular}
\end{table}

\subsubsection{Parents' Experience With \quotes{Kids} and \quotes{Family} Stores}
We asked parents if they have heard about the \quotes{Kids} app category on Apple's app store or the \quotes{Family} category on the Google Play store, which provide a content rating for apps into different age categories, such as under 5, between 6 and 8, or between 9 and 11 years. We find that 77 (64.71\%) Saudi parents have heard about the aforementioned \quotes{Kids} or \quotes{Family} categories in app stores. Out of those who have heard about at least one of these store categories, a total of 87.01\% have tried to use them at some level, where 29.87\% use them \quotes{all the time}, 36.36\% use them \quotes{sometimes}, and 10.39\% use them \quotes{rarely}. See~\autoref{tab:experience} for detailed results. Out of those who have tried to use at least one of these stores at any level, a total of 80.60\% either \quotes{strongly agree} or \quotes{somewhat agree} that using the \quotes{Kids} or \quotes{Family} categories in app stores reduces their privacy concerns about what their children use on the device. See~\autoref{tab:usage} for the full results. \par 

\begin{table}[!t]
\centering
\caption{Parents who used the \quotes{Kids} or the \quotes{Family} app store categories, or use parental control apps for their children.}
\label{tab:experience}
\resizebox{\columnwidth}{!}{
\begin{tabular}{ll@{\hspace{5pt}}rl@{\hspace{5pt}}r}
\toprule	
\multirow{2}{*}{Frequency} & \multicolumn{4}{c}{Parents}\\
\cline{2-5}
 & \multicolumn{2}{c}{Use \quotes{Kids}/\quotes{Family}} & \multicolumn{2}{c}{Use Parental Control} \\
\hline 
All the time & 23 & (29.87\%) & 16 & (15.53\%)\\
Sometimes & 28 & (36.36\%) & 29 & (28.16\%)\\
Rarely & 8 & (10.39\%) & 8 & (7.77\%)\\
Tried, but found them unhelpful & 5 & (6.49\%) & 9 & (8.74\%)\\
Used to, but no longer find them helpful & 3 & (3.90\%) & 4 & (3.88\%)\\
Never used them & 10 & (12.99\%) & 37 & (35.92\%) \\
\bottomrule
\end{tabular}
}
\end{table}
\begin{table}[!t]
\centering
\caption{Parents' level of agreement towards Q. 43 statement: \quotes{I feel that Apple's \quotes{Kids} or Google's \quotes{Family} categories reduce my privacy concerns about what my child uses on the device}.}
\label{tab:usage}
\begin{tabular}{ll@{\hspace{5pt}}r}
\toprule
Level of agreement & \multicolumn{2}{c}{Parents} \\
\hline 
Strongly agree & 12 & (17.91\%) \\
Somewhat agree & 42 & (62.69\%) \\
Neither agree nor disagree (neutral) & 12 & (17.91\%) \\
Somewhat disagree & 1 & (1.49\%) \\
Strongly disagree & 0 & (0\%) \\
\bottomrule
\end{tabular}
\end{table}

Comparing Saudi parents' awareness and use of \quotes{Kids} or \quotes{Family} categories in app stores with their Western and Chinese counterparts in~\cite{zhao18,wang19}, contrary to Saudi parents, we find that most Western parents (61\%) have not heard about the \quotes{Kids} or \quotes{Family} app store categories. However, similar to Saudi parents, of those Western parents who have heard about them, 79\% use them either \quotes{all the time} or \quotes{sometimes}. Of those Western parents who used them, 66.5\% \quotes{strongly agree} or \quotes{somewhat agree} that they reduce their privacy concerns about what their child uses on the device.

\subsubsection{Parents' Experience With Parental Control Apps}
We asked parents if they have heard about \quotes{parental control} apps, which allow parents to control their children's tablet or smartphone devices, such as controlling the number of hours their children can spend on the device, creating a list of disallowed apps, and a list of disallowed websites. We find that 103 (86.55\%) of Saudi parents have heard about these apps. Out of those who have heard about them, a total of 64.08\% have tried them at some level, and a total of 12.62\% have tried them or have been using them \quotes{but find them unhelpful}, while 35.92\% \quotes{never used them}. We inspected the children's ages for the parents who reported that they use parental control apps frequently, i.e. either \quotes{all the time} or \quotes{sometimes}. Out of 45 parents who use these apps frequently, we find the highest number of parents who use these apps is for parents of 7-year-old children, where 11 (24.44\%) parents of 7-year-old children reported that they use parental control apps, followed by 9 (20\%) parents of 12-year-old children, then equally 6 (13.33\%) parents for each of the remaining ages (6, 9, 10, 11)-year-old children reported that they use these apps, while only one parent (2.22\%) for an 8-year-old child uses these apps frequently. One possible explanation for the strikes of parental control apps shown by parents of 7 and 12-year-old children is that children at the age of 7 and 12 are in turning points (from kindergarten to primary school at the age of 7, or from primary school to intermediate school at the age of 12), since children in Saudi Arabia are ideally admitted to primary school at the age of 6 or 7 years. It is also surprising to see that over 35\% of the parents who have heard about parental control apps state that they never used them. They might have some profound reasons such as a lack of technical skills, or a negative reputation about these apps (e.g. difficult to use). See~\autoref{tab:experience} for the full results. Finally,~\autoref{tab:parental_control} lists the most used parental control apps that are reported by those who use them.

\begin{table}[!t]
	\centering
	\caption{The most commonly used parental control apps.}
	\label{tab:parental_control}
	\begin{tabular}{lc}
		\toprule
		App name & Parents \\
		\hline 
		screen time & 6 \\
		ourpact 	& 4 \\
		family link	& 4 \\
		qustodio 	& 2 \\
		familytime	& 1 \\
		safekids	& 1 \\
		net nanny	& 1 \\
		netflix parental control & 1 \\ 	 
		\bottomrule
	\end{tabular}
\end{table}

We could not compare this section's result against the results of Western and Chinese parents as it does not exist in the relevant studies in~\cite{zhao18,wang19}.

\section{Digital Divide} \label{sec:divide}
In this section, we examine whether there are statistically significant differences between high versus low socioeconomic classes in Saudi society in terms of privacy practices and concerns about their children's smart device applications. We denote these differences by the term \quotes{digital divide}.

\subsection{Analysis Methodology}
We considered Saudi society from three socioeconomic perspectives: parents' education, technical background, and income. For each perspective, we divided the responses into two classes: high and low. We defined the high and low classes of each socioeconomic perspective in~\autoref{tab:levels}. It should be noted that the total number of responses in each class of these perspectives is a partial set of the overall responses. That is, we classified a family in either a high or low class only if \quotes{both} parents' demographics met the high or low definition. That is, in our analysis, we did not count responses where only \quotes{one} of the parents met the high or low class definition. We also did not count responses where the answer is not firm, such as \quotes{Not applicable}, \quotes{I do not know}, and \quotes{Other}. Those excluded responses from high or low classes are also excluded from the \quotes{total} counts. The reasons for these exclusions are to have firm definitions and to exclude noise, hence gaining more confidence in the results. \par 

In terms of our definitions that are listed in~\autoref{tab:levels}, first, for the education perspective, we defined high class education as both parents have a postgraduate (Master's or Doctorate) degrees. We defined low class education as both parents have a Bachelor's degree at most. Second, for the technical background perspective, we adopted a universally accepted indicator of what constitutes a technical background. We defined high technical background as both parents work or have a degree in CS, IS, IT, or CE. We defined low technical background as both parents neither work nor have a degree in CS, IS, IT, or CE. Third, for the income perspective, we based our definitions of high versus low income on a national study that defined the average income of what is called the \quotes{sufficiency line}~\cite{aldamegh14} (more details about the income classes that we defined are provided in~\autoref{sec:economic}). \par

\begin{table*}[!t]
\centering 
\caption{Definitions of high and low (education, technical background, and income) which we use to examine the digital divide.}
\label{tab:levels}
\resizebox{\textwidth}{!}{
\begin{tabular}{llll}
\toprule
Perspective & High class definition & low class definition & Q.\\
\midrule
Education & Both parents have either a Master's or a Doctorate as their highest degree & Both parents have at most a Bachelor degree as their highest degree & Q. 8 \& Q. 10\\
\hline
Technical background & Both parents work or have a degree in CS, IS, IT or CE & Both parents neither work nor have a degree in CS, IS, IT or CE & Q. 9 \& Q. 12\\
\hline
Income & The household monthly income is more than \num{34000} SAR & The household monthly income is at most \num{34000} SAR & Q. 13\\
\bottomrule
\end{tabular}
}
\end{table*}
After we defined the classes and extracted the responses that meet the definition of each class, we then examined the responses of each class (e.g. high education responses versus low education responses) against certain privacy-related properties. These properties represent parents' privacy practices and concerns, and are extracted from the survey questions. That is, from the respondent's answer, we can identify if it meets the property or not.~\autoref{tab:properties} shows the properties we examined the socioeconomic classes against, and the conditions under which the properties are satisfied (i.e. met), along with the survey questions they are extracted from.\par 

\begin{table*}[!t]
\centering 
\caption{Definitions of high and low (education, technical background, and income) which we use to examine the digital divide.}
\label{tab:properties}
\begin{tabularx}{\linewidth}{lLl} 
\toprule
Property & Satisfied (i.e. met) if: & Q. \\
\midrule
Use password  & Parents have a password to log in to the device or to install apps for their children's devices & Q. 24\\
\hline
Used paid app instead of free for privacy reasons  & Parents have chosen a paid app of an otherwise free app to reduce privacy risks  & Q. 27 \\
\hline
Think about privacy frequently & Parents think about privacy \quotes{at least once a month}, \quotes{at least once a week}, or \quotes{at least once a day} & Q. 32 \\
\hline
Talk about privacy frequently & Parents talk about privacy \quotes{at least once a month}, \quotes{at least once a week}, or \quotes{at least once a day} & Q. 34 \\
\hline
High level of concern about camera \& microphone (\textit{before}) & Parents are \quotes{very concerned} or \quotes{somewhat concerned} regarding apps requesting camera and microphone access in their children's devices \textit{before} they are informed about possible privacy implications & Q. 35 \\
\hline
High level of concern about location (\textit{before}) & Parents are \quotes{very concerned} or \quotes{somewhat concerned} regarding apps requesting location access in their children's devices \textit{before} they are informed about possible privacy implications & Q. 36 \\
\hline
High level of concern about camera \& microphone (\textit{after}) & Parents are \quotes{very concerned} or \quotes{somewhat concerned} regarding apps requesting camera and microphone access in their children's devices \textit{after} they are informed about possible privacy implications & Q. 37 \\
\hline
High level of concern about location (\textit{after}) & Parents are \quotes{very concerned} or \quotes{somewhat concerned} regarding apps requesting location access in their children's devices \textit{after} they are informed about possible privacy implications  & Q. 38 \\
\hline
Aware about \quotes{Kids/Family} apps & Parents have heard about the \quotes{Kids} or \quotes{Family} app store categories & Q. 41 \\
\hline
Use \quotes{Kids/Family} apps frequently & Parents use the \quotes{Kids} or \quotes{Family} app store categories \quotes{all the time} or \quotes{sometimes} & Q. 42\\
\hline
Aware about privacy settings & Parents have heard about privacy permission settings of an app & Q. 44\\
\hline
Check privacy settings frequently & Parents check privacy permission settings for their children's devices \quotes{at least once a month}, \quotes{at least once a week}, or \quotes{at least once a day} & Q. 45\\
\hline
Aware about parental control apps & Parents have heard about parental control apps & Q. 48\\
\hline
Use parental control apps frequently & Parents use parental control apps \quotes{all the time} or \quotes{sometimes} & Q. 49 \\
\bottomrule
\end{tabularx}
\end{table*}
After counting the number of low and high responses that satisfied the properties defined in~\autoref{tab:properties}, we conducted the significant differences test using the Chi-square test of independence or the Fisher--Irwin test (the latter is used only if the expected frequency is less than 5)~\cite{campbell07}, with a level of significance alpha~=~0.05.

\subsection{Is there a digital divide?}
After computing the significance tests, out of 42 tests, we find significant differences (digital divide) between high versus low classes in 7 tests only. This is a positive trend overall. However, it is important to work on bridging these gaps. In what follows, we summarise those areas of concern, which are summarised in~\autoref{tab:divide1},~\autoref{tab:divide2}, and~\autoref{tab:divide3}. The results show that:

\begin{enumerate}
\item \textbf{Parents' education is related to awareness about \quotes{Kids} and \quotes{Family} app store categories.} That is, 85.19\% high education parents are aware about the \quotes{Kids} or \quotes{Family} app store categories, compared to 58.82\% low education parents who do so. This difference is significant ($\chi^2(1) = 5.63$, $p = 0.02 (< 0.05)$, $n = 78$).

\item \textbf{Out of those who are aware of privacy settings, parents' education, technical background, and income are related to checking privacy settings in their children's devices frequently.} That is, out of those who are aware about privacy settings, 61.76\% low education parents, 62\% low technical background parents, and 54.10\% low income parents, check privacy settings frequently (i.e. \quotes{at least once a month}, \quotes{at least once a week}, or \quotes{at least once a day}), compared to 21.74\% high education parents, 15.38\% high technical background parents, and 20\% high income parents. These differences are significant. For education: ($\chi^2(1) = 8.86$, $p = 0.00 (< 0.05)$, $n = 57$), for technical background: ($\chi^2(1) = 8.99$, $p = 0.00 (< 0.05)$, $n = 63$), and for income: ($\chi^2(1) = 7.06$, $p = 0.01 (< 0.05)$, $n = 81$). These results suggest that parents from low socioeconomic classes show higher levels of concern about their children's privacy in this aspect and are more responsive than parents from high socioeconomic classes. That is, once they become aware of privacy settings, there are more parents from low socioeconomic classes who check their children's privacy settings frequently than parents from high socioeconomic classes. 

\item \textbf{Parents' technical background is related to using passwords in their children's devices.} That is, 93.75\% high technical background parents use passwords compared to 62.32\% low technical background parents. This difference is significant ($\chi^2(1) = 5.92$, $p = 0.01 (< 0.05)$, $n = 85$).

\item \textbf{Parents' income is related to thinking about their children's privacy frequently.} That is, 96.34\% low income parents think about their children's privacy frequently (i.e. \quotes{at least once a month}) compared to 80.77\% high income parents. This difference is significant\footnote{There is no $\chi^2$ value as we used the Fisher--Irwin test} ($p = 0.02 (< 0.05)$, $n = 108$). Moreover, out of the 79 low income parents who think about their children's privacy frequently (i.e. \quotes{at least once a month}, \quotes{at least once a week}, or \quotes{at least once a day}), 50 (63.29\%) think about it \quotes{at least once a day}. 

\item \textbf{Out of those who are aware of the \quotes{Kids} or \quotes{Family} app store categories, parents' income is related to using them frequently.} That is, 88.24\% high income parents use the \quotes{Kids} or \quotes{Family} app store categories frequently (i.e. \quotes{all the time} or \quotes{sometimes}), compared to 57.41\% of low income parents. This difference is significant ($\chi^2(1) = 5.39$, $p = 0.02 (< 0.05)$, $n = 71$).

\end{enumerate}

\begin{table*}[!t]
\centering
\caption{The statistical difference between high and low education against the listed properties. \quotes{No.} denotes the number of (low $\vert$ high) education parents who satisfy the property listed in the column labeled \quotes{Property}, \quotes{Total} denotes the total number of (low $\vert$ high) education parents, $p$ denotes the p-value, $\varphi$ denotes the effect size, $n$ denotes the total sample, $\chi^2(1)$ denotes chi-square test statistic with one degree of freedom. Note that the total numbers (\quotes{Total}) of (low $\vert$ high) education parents in the properties number 10, 12, and 14 are derived from the numbers (\quotes{No.}) of the previous properties (9, 11, and 13) respectively. This is because the related questions for the properties 10, 12, and 14 are only shown to participants who answered \quotes{yes} for the awareness-related questions which are related to properties 9, 11, and 13 respectively.}
\label{tab:divide1}
\resizebox{\textwidth}{!}{
\begin{tabular}{clllrrlrrll}
\toprule
\rowcolor{gray!50}
\multicolumn{11}{c}{Education}\\
\midrule
Serial No. & Property        & & \multicolumn{2}{c}{High Education} & & \multicolumn{2}{c}{Low Education} & & \multirow{2}{*}{$\chi^2(1)$}  & \multirow{2}{*}{Fisher}   \\
\cline{4-5}\cline{7-8}
 &  & & No./Total & \%  & & No./Total & \%  & & &\\ 
\hline
 9 & Aware about \quotes{Kids/Family} apps
  & &   23/27  & (85.19\%) & &   30/51  & (58.82\%)  & &    
\begin{tabular}{@{}l@{}} $\chi^2(1)$ = 5.63 \\ $p$ = 0.02 \\ $\varphi$ = 0.27 \\ $n$ = 78 \end{tabular} & -  \\ 
\hline
 12 &  Check privacy settings frequently
  & &   5/23  & (21.74\%) & &  21/34 & (61.76\%)   & &    
\begin{tabular}{@{}l@{}} $\chi^2(1)$ = 8.86 \\ $p$ = 0.00 \\ $\varphi$ = 0.39 \\ $n$ = 57 \end{tabular} & -  \\  
\bottomrule
\end{tabular}
} 
\end{table*}

\begin{table*}[!t]
\centering
\caption{The statistical difference between high and low technical background level against the listed properties. \quotes{No.} denotes the number of (low $\vert$ high) technical background parents who satisfy the property listed in the column labeled \quotes{Property}, \quotes{Total} denotes the total number of (low $\vert$ high) technical background parents, $p$ denotes the p-value, $\varphi$ denotes the effect size, $n$ denotes the total sample, $\chi^2(1)$ denotes chi-square test statistic with one degree of freedom. Note that the total numbers (\quotes{Total}) of (low $\vert$ high) technical background parents in the properties number 10, 12, and 14 are derived from the numbers (\quotes{No.}) of the previous properties (9, 11, and 13) respectively. This is because the related questions for the properties 10, 12, and 14 are only shown to participants who answered \quotes{yes} for the awareness-related questions which are related to properties 9, 11, and 13 respectively.}
\label{tab:divide2}
\resizebox{\textwidth}{!}{
\begin{tabular}{clllrrlrrll}
\toprule
\rowcolor{gray!50}
\multicolumn{11}{c}{Technical Background}\\
\midrule
Serial No. & Property        & & \multicolumn{2}{c}{High Technical Background} & & \multicolumn{2}{c}{Low Technical Background} & & \multirow{2}{*}{$\chi^2(1)$} & \multirow{2}{*}{Fisher}  \\
\cline{4-5}\cline{7-8}
 &  & & No./Total & \%  & & No./Total & \%  & & & \\ 
\hline
1 & Use password             
   & &  15/16 & (93.75\%)  & &   43/69 & (62.32\%) & &  
\begin{tabular}{@{}l@{}} $\chi^2(1)$ = 5.92 \\ $p$ = 0.01 \\ $\varphi$ = 0.26 \\ $n$ = 85 \end{tabular} & -  \\ 
\hline
12 & Check privacy settings frequently
  & &   2/13  & (15.38\%)  & &  31/50 & (62.00\%)   & &  
\begin{tabular}{@{}l@{}} $\chi^2(1)$ = 8.99 \\ $p$ = 0.00 \\ $\varphi$ = 0.38  \\ $n$ = 63 \end{tabular} & -  \\  
\bottomrule
\end{tabular}
} 
\end{table*}


\begin{table*}[!t]
\centering
\caption{The statistical difference between high and low income against the listed properties. \quotes{No.} denotes the number of (low $\vert$ high) income parents who satisfy the property listed in the column labeled \quotes{Property}, \quotes{Total} denotes the total number of (low $\vert$ high) income parents, $p$ denotes the p-value, $\varphi$ denotes the effect size, $n$ denotes the total sample, $\chi^2(1)$ denotes chi-square test statistic with one degree of freedom. Note that the total numbers (\quotes{Total}) numbers of (low $\vert$ high) income parents in the properties number 10, 12, and 14 are derived from the numbers (\quotes{No.}) of the previous properties (9, 11, and 13) respectively. This is because the related questions for the properties 10, 12, and 14 are only shown to participants who answered \quotes{yes} for the awareness-related questions which are related to properties 9, 11, and 13 respectively.}
\label{tab:divide3}
\resizebox{\textwidth}{!}{
\begin{tabular}{clllrrlrrll}
\toprule
\rowcolor{gray!50}
\multicolumn{11}{c}{Income}\\
\midrule
Serial No. & Property        & & \multicolumn{2}{c}{High Income} & & \multicolumn{2}{c}{Low Income} & & \multirow{2}{*}{$\chi^2(1)$} & \multirow{2}{*}{Fisher} \\
\cline{4-5}\cline{7-8}
  & & & No./Total & \%  & & No./Total & \%  & & &\\ 
\hline
3 & Think about privacy frequently
& &  21/26 & (80.77\%)  & &  79/82 & (96.34\%)   & &
- & \begin{tabular}{@{}l@{}} $p$ = 0.02 \\ $\varphi$ = 0.25 \\ $n$ = 108 \end{tabular} \\ 
\hline
10 & Use \quotes{Kids/Family} apps frequently
& &  15/17  & (88.24\%)  & &   31/54  & (57.41\%)   & &   
\begin{tabular}{@{}l@{}} $\chi^2(1)$ = 5.39 \\ $p$ = 0.02 \\ $\varphi$ = 0.28 \\ $n$ = 71 \end{tabular} & - \\  
\hline
12 & Check privacy settings frequently
& &   4/20  & (20.00\%)  & & 33/61 & (54.10\%)   & & 
\begin{tabular}{@{}l@{}} $\chi^2(1)$ = 7.06 \\ $p$ = 0.01 \\ $\varphi$ = 0.30 \\ $n$ = 81 \end{tabular} & - \\  
\bottomrule
\end{tabular}
} 
\end{table*}

\section{Recommendations}\label{sec:recomm}
Having discussed the results, we now list the recommendations for parents, professionals (such as researchers and developers), regulators, and policy makers. 
\subsection{For Parents}
We provide the following recommendations for parents, guardians, teachers, or whoever is in a position of care for children's safety online:

\begin{enumerate}
\item \textbf{Check apps' age appropriateness for children.}
Most app stores (e.g. the Google Play store and Apple's App store) provide additional information about the apps to check an app's age appropriateness for children, and what permissions the app asks for. For example, the Google Play store has the \quotes{Content rating} and \quotes{Permission} sections under the \quotes{Additional Information} section for each app. Similarly, Apple's App store has the \quotes{age}, \quotes{data used to track you}, and \quotes{data linked to you} sections. Our results show that half of the top 10 most used apps by Saudi children who are aged between 6 to 12 years are rated for 12+ years and require parental guidance. While these rating systems do not seem perfect and parents need to make their own judgment (e.g. \quotes{MS Teams} is listed as suitable for children aged 3+ years in the Google Play store), they can be used as an initial assessment tool. After checking the app's \quotes{Content rating} or \quotes{Age}, parents may play around with the apps themselves, to further assess the app for their young children (6 to 12-year-old as in our study).

\item \textbf{Get involved with children during the apps' installation stage before the app is installed by the child.} Our results suggest that parents' involvement in installing their children's apps is related to reduced uninstallation incidents by parents for their children's apps, which are mostly negatively received by children.

\item \textbf{When selecting apps for children, it is not enough to pay attention to the apps' content only.} The apps' requested permissions and data collection are equally important to the apps' content. We recommend thinking about data from two ways: what the children receive (content) and what the children provide or the app accesses (privacy). Apps might collect personal data such as location, microphone, camera, contacts, or device ID data. Parents need to be aware of the apps' terms and conditions and the permissions granted to an app. The terms and conditions tell parents how the child's data will be used including whether or not it will be shared with other parties and for what purposes. The data an app collects can be checked either before installing the app or after it is installed. Before installing an app, most app stores provide a list of permissions that an app requests in an easy-to-read format such as the apps' \quotes{Additional Information} section in the Google Play store. This information helps parents make an informed decision about whether or not to share certain types of personal data. Parents can also check and tune an app's permissions after the app is installed from the privacy settings of the device by choosing the service, e.g. camera or location, and check the list of apps that have access to a particular service. Parents can deny one or more apps from accessing that service (e.g. location), and make an informed decision based on some factors, e.g. whether the permissions the app requests are proportionate to the app's function or not. However, some apps may not function without access to some permissions. Making an informed decision is the key point here. Our results show that, when installing an app for their children, parents are more concerned about the app's content and what the app does than the app's access to sensitive personal information or the app's requested permissions.

\item \textbf{Set passwords (or an alternative access control mechanism) to control both login to the device and installation of free and paid apps}. The device's login password protects against outsiders, while the apps' installation password protects against children's installation or purchasing of apps without their parents' consent. Our results show that around a third of the parents do not set passwords to control login to the device nor to control apps installation on their children's devices.

\item \textbf{Make use of the \quotes{Kids} and \quotes{Family} app store categories.} While they are not a panacea, they reduce parents' privacy concerns as reported. 
\end{enumerate}

\subsection{For Professionals}
We provide the following recommendations for professionals such as researchers and developers:
\begin{enumerate}
\item \textbf{Raise awareness about the aforementioned issues.} All the aforementioned recommendations to parents require some level of technical background which may not be available to everyone. Professionals such as researchers and developers need to invest in privacy education for everyone, and particularly for parents towards their children's privacy online. Our results show that parents seem very eager to learn about privacy, where 100\% of those who have not heard about privacy permissions for an app reported that they are interested to learn more about them. Towards this direction, several studies in the literature, such as~\cite{zhao19,kennedy16,kumar17,dowthwaite20,urquiola19,urquiola17,fitton19}, focused on building frameworks, and understanding children's, youths', and parents' perceptions of privacy risks online and the used or perceived techniques to overcome them. Moreover, we recommend professionals raise awareness about data minimisation by design (or more generally \quotes{privacy by design} principles~\cite{cavoukian11}) among the scientific and developer communities.

\item \textbf{Use \quotes{effective} awareness. Awareness alone is not enough.} Awareness needs to be effectively designed using a language that is comprehended by everyone. Our results show that stating privacy implications (what can go wrong) in a scenario-based format has significantly increased parents' concern about privacy. This is a useful tip that our results provide for awareness content delivery. The literature proposed several attempts that aimed to provide effective awareness tools to children, such as the physical-digital interactive book for youth~\cite{yap20}, the \quotes{LifeMosaic} personal informatics tool for youth~\cite{potapov20}, and games~\cite{kumar18}.
\end{enumerate}

\subsection{For Regulators and Policy Makers}
We provide the following recommendations for regulators and policy makers:
\begin{enumerate}
\item \textbf{Set guidelines and regulations that define transparent and fair data collection.} For example, in Europe there is the GDPR~\cite{gdpr21}, in the US, there is the Children's Online Privacy Protection Act (COPPA)~\cite{ftc98}, and more recently in the UK, the \quotes{Age Appropriate Design Code} was introduced~\cite{ico21}. COPPA sets certain requirements for online services directed to children. With the rapid uptake of digital technologies and the lack of data collection regulations, non-regulated data collection by third parties can be harmful in the long run. It may normalise sharing personal data with third parties. Having said that, regulations such as GDPR and COPPA have their own technical and non-technical issues, such as the challenges in monitoring compliance, which may need non-conventional tools. For example, Reyes et al.~\cite{reyes18} found many apps available in the US are violating COPPA. Nevertheless, we can learn from them and work towards improving their issues.
\end{enumerate}

\section{Limitations} \label{sec:limit}
In this section, we list some limitations in our work. Most of these limitations have been discussed and justified earlier in the text. However, we list them here for summarisation. First, despite our efforts to distribute our survey to a diverse population of both genders and various socioeconomic and educational backgrounds, our respondents are biased towards highly educated, females, who are from the Western region of Saudi Arabia. Second, we did not share the survey link over \quotes{public} social media accounts. This might have limited the number of responses we could receive. However, as stated in the methodology~\autoref{sec:deployment_actual}, we did this to avoid incorrect or duplicate responses from irresponsible users, or possibly from automated software bots. Third, we discussed some aspects of our results with results from two previous surveys conducted on Western (mostly from the UK) and Chinese parents~\cite{zhao18,wang19}. However, it should be noted that the comparisons are high-level and are not meant to be direct comparisons. There are slight differences in the demographics and recruitment methodologies between the three studies, which are explained in detail in the analysis in~\autoref{sec:analysis}. Having said that, our work is independent of~\cite{zhao18,wang19}, and our results stand on their own regardless of~\cite{zhao18,wang19} results. The comparisons are meant to foster discussions, and identify patterns and relationships, which may inspire future work.  

\section{Conclusion} \label{sec:conclusion}
In this paper, we conducted a survey to investigate Saudi parents' concerns about their children's privacy when using smart device apps, and what are they doing to mediate these concerns. To this end, we analysed 119 responses and identified parents' practices and concerns. We found that Saudi parents expressed a high level of concern regarding their children's privacy when using smart device apps. However, they expressed higher concerns about apps' content than privacy issues such as apps' requests to access sensitive data. Furthermore, parents' concerns are not in line with most of the children's installed apps, which contain apps inappropriate for their age, require parental guidance, and request access to sensitive data such as location. We also discuss Saudi parents' privacy practices and concerns with those outlined by Western and Chinese parents in previous reports. This discussion revealed interesting patterns and allowed us to draw new relationships. For example, Saudi and Chinese children have a higher level of autonomy in installing apps on their devices than Western children who mostly request apps installation from their parents. However, there are more Saudi parents who end up uninstalling apps from their children's devices due to privacy concerns than Western parents. Moreover, Saudi and Western parents show higher levels of privacy concerns than Chinese parents. We then tested whether there are significant differences between high vs. low socioeconomic classes in terms of privacy practices and concerns (we denote these differences by \quotes{digital divide}). We looked at the society from three perspectives: parents' education, technical background, and income. Out of 42 tests, we find significant differences in 7 tests only. Finally, we made recommendations to improve children's privacy online, which require efforts from multiple parties. The results of this study help us better understand our situation to identify areas of improvements, set recommendations, and develop the right tools towards creating a safer online world for our children.

\section{Acknowledgments} \label{sec:ack}
We thank the pilot survey participants for their time and feedback. We thank Wafaa Alashwali for pointing the inflation rate we mentioned in~\autoref{sec:economic} to us, and for help in calculating it. We thank Jun Zhao from the University of Oxford for providing us with the survey used in~\cite{zhao18} and for feedback. We thank all the survey participants for their time and valuable answers. Finally we thank everyone who helped us in distributing the survey. 


\bibliographystyle{IEEEtran}
\bibliography{refs}

\begin{thebibliography}{10}
\providecommand{\url}[1]{#1}
\csname url@samestyle\endcsname
\providecommand{\newblock}{\relax}
\providecommand{\bibinfo}[2]{#2}
\providecommand{\BIBentrySTDinterwordspacing}{\spaceskip=0pt\relax}
\providecommand{\BIBentryALTinterwordstretchfactor}{4}
\providecommand{\BIBentryALTinterwordspacing}{\spaceskip=\fontdimen2\font plus
\BIBentryALTinterwordstretchfactor\fontdimen3\font minus
  \fontdimen4\font\relax}
\providecommand{\BIBforeignlanguage}[2]{{%
\expandafter\ifx\csname l@#1\endcsname\relax
\typeout{** WARNING: IEEEtran.bst: No hyphenation pattern has been}%
\typeout{** loaded for the language `#1'. Using the pattern for}%
\typeout{** the default language instead.}%
\else
\language=\csname l@#1\endcsname
\fi
#2}}
\providecommand{\BIBdecl}{\relax}
\BIBdecl

\bibitem{ofcom20}
\BIBentryALTinterwordspacing
Ofcom, ``Children and {P}arents: {M}edia {U}se and {A}ttitudes {R}eport 2019,''
  2020, accessed on: Jan. 12, 2021. [Online]. Available:
  \url{https://www.ofcom.org.uk/__data/assets/pdf_file/0023/190616/children-media-use-attitudes-2019-report.pdf}
\BIBentrySTDinterwordspacing

\bibitem{commonsense19}
\BIBentryALTinterwordspacing
{Common Sense}, ``The {C}ommon {S}ense {C}ensus: {M}edia {U}se by {T}weens and
  {T}eens,'' 2019, accessed on: Jan. 12, 2021. [Online]. Available:
  \url{https://www.commonsensemedia.org/sites/default/files/uploads/research/2019-census-8-to-18-full-report-updated.pdf}
\BIBentrySTDinterwordspacing

\bibitem{norton18}
\BIBentryALTinterwordspacing
Norton, ``Norton's {M}y {F}irst {D}evice {R}eport,'' 2018, accessed on: Jan.
  13, 2020. [Online]. Available:
  \url{https://now.symassets.com/content/dam/norton/global/pdfs/reports/Norton_My_First_Device_Report_Oct_2018_Final.pdf}
\BIBentrySTDinterwordspacing

\bibitem{gas_yearbook_19}
\BIBentryALTinterwordspacing
{General Authority for Statistics (GAS)}, ``Statistical {Y}earbook of 2019 |
  {I}ssue {N}umber: 55,'' 2019, accessed on: Apr. 15, 2021. [Online].
  Available: \url{https://www.stats.gov.sa/en/1006}
\BIBentrySTDinterwordspacing

\bibitem{zhao18}
\BIBentryALTinterwordspacing
J.~Zhao, U.~Lyngs, and N.~Shadbolt, ``What {P}rivacy {C}oncerns {D}o {P}arents
  {H}ave {A}bout {C}hildren's {M}obile {A}pps, and {H}ow {C}an {T}hey {S}tay
  {SHARP}?'' University of Oxford, KOALA Project Report 1, 2018, accessed on:
  Dec. 18, 2020. [Online]. Available:
  \url{https://arxiv.org/pdf/1809.10841.pdf}
\BIBentrySTDinterwordspacing

\bibitem{wang19}
\BIBentryALTinterwordspacing
G.~Wang, J.~Zhao, and N.~Shadbolt, ``What {C}oncerns {D}o {C}hinese {P}arents
  {H}ave {A}bout {T}heir {C}hildren's {D}igital {A}doption and {H}ow to
  {B}etter {S}upport {T}hem ?'' University of Oxford, KOALA Project Report 3.5,
  2019, accessed on: Dec. 18, 2020. [Online]. Available:
  \url{https://arxiv.org/pdf/1906.11123.pdf}
\BIBentrySTDinterwordspacing

\bibitem{wikipedia21_sa}
\BIBentryALTinterwordspacing
Wikipedia, ``Saudi {A}rabia,'' 2021, accessed on: Mar. 11, 2021. [Online].
  Available: \url{https://en.wikipedia.org/wiki/Saudi_Arabia}
\BIBentrySTDinterwordspacing

\bibitem{gas19}
\BIBentryALTinterwordspacing
{General Authority for Statistics (GAS)}, ``Population {E}stimates,'' 2019,
  accessed on: Jan. 30, 2021. [Online]. Available:
  \url{https://www.stats.gov.sa/en/43}
\BIBentrySTDinterwordspacing

\bibitem{opec20}
\BIBentryALTinterwordspacing
{Organization of the Petroleum Exporting Countries (OPEC)}, ``{OPEC} : {S}audi
  {A}rabia,'' 2021, accessed on: Jan. 12, 2021. [Online]. Available:
  \url{https://www.opec.org/opec_web/en/about_us/169.htm}
\BIBentrySTDinterwordspacing

\bibitem{gassem19}
\BIBentryALTinterwordspacing
L.~B. Gassem, ``38 {P}ercent of {P}arents in {M}iddle {E}ast {S}truggle to
  {P}rotect their {C}hildren {O}nline,'' 2019, accessed on: Dec. 17, 2021.
  [Online]. Available: \url{https://www.arabnews.com/node/1449636/saudi-arabia}
\BIBentrySTDinterwordspacing

\bibitem{isaak18}
J.~Isaak and M.~J. Hanna, ``User {D}ata {P}rivacy: {F}acebook, {C}ambridge
  {A}nalytica, and {P}rivacy {P}rotection,'' \emph{Computer}, vol.~51, no.~8,
  pp. 56--59, 2018.

\bibitem{cadwalladr18}
\BIBentryALTinterwordspacing
C.~Cadwalladr and E.~Graham-Harrison, ``Revealed: 50 {M}illion {F}acebook
  {P}rofiles {H}arvested for {C}ambridge {A}nalytica in {M}ajor {D}ata
  {b}reach,'' 2018, accessed on: Dec. 3, 2021. [Online]. Available:
  \url{https://www.theguardian.com/news/2018/mar/17/cambridge-analytica-facebook-influence-us-election}
\BIBentrySTDinterwordspacing

\bibitem{rosenberg18}
\BIBentryALTinterwordspacing
M.~Rosenberg, N.~Confessore, and C.~Cadwalladr, ``How {T}rump {C}onsultants
  {E}xploited the {F}acebook {D}ata of {M}illions,'' 2018, accessed on: Dec. 3,
  2021. [Online]. Available:
  \url{https://www.nytimes.com/2018/03/17/us/politics/cambridge-analytica-trump-campaign.html}
\BIBentrySTDinterwordspacing

\bibitem{confessore18}
\BIBentryALTinterwordspacing
N.~Confessore, ``Cambridge {A}nalytica and {F}acebook: {T}he {S}candal and the
  {F}allout {S}o {F}ar,'' 2018, accessed on: Dec. 3, 2021. [Online]. Available:
  \url{https://www.nytimes.com/2018/04/04/us/politics/cambridge-analytica-scandal-fallout.html}
\BIBentrySTDinterwordspacing

\bibitem{kennedy19}
A.-M. Kennedy, K.~Jones, and J.~Williams, ``Children as {V}ulnerable
  {C}onsumers in {O}nline {E}nvironments,'' \emph{Journal of Consumer Affairs},
  vol.~53, no.~4, pp. 1478--1506, 2019.

\bibitem{livingstone14}
S.~Livingstone, L.~Kirwil, C.~Ponte, and E.~Staksrud, ``In {T}heir {O}wn
  {W}ords: {W}hat {B}others {C}hildren {O}nline?'' \emph{European Journal of
  Communication}, vol.~29, no.~3, pp. 271--288, 2014.

\bibitem{gazette20}
\BIBentryALTinterwordspacing
{Saudi Gazette}, ``Two {I}nitiatives by {MBS} to {S}erve {G}lobal {C}yber
  {S}ecurity {A}nnounced,'' 2020, accessed on: Dec. 24, 2020. [Online].
  Available: \url{https://saudigazette.com.sa/article/588277}
\BIBentrySTDinterwordspacing

\bibitem{hasebrink09}
\BIBentryALTinterwordspacing
U.~Hasebrink, S.~Livingstone, L.~Haddon, and K.~\'{O}lafsson, ``{EU} {K}ids
  {O}nline: {C}omparing {C}hildren's {O}nline {O}pportunities and {R}isks
  {A}cross {E}urope,'' 2009, accessed on: May 3, 2021. [Online]. Available:
  \url{http://eprints.lse.ac.uk/24368/1/D3.2_Report-Cross_national_comparisons-2nd-edition.pdf}
\BIBentrySTDinterwordspacing

\bibitem{dedkova20}
L.~Dedkova, D.~Smahel, and M.~Just, ``Digital {S}ecurity in {F}amilies: the
  {S}ources of {I}nformation {R}elate to the {A}ctive {M}ediation of {I}nternet
  {S}afety and {P}arental {I}nternet {S}kills,'' \emph{Behaviour \& Information
  Technology}, pp. 1--13, 2020.

\bibitem{regents21}
\BIBentryALTinterwordspacing
{The Regents of the University of Michigan}, ``Sharing {T}oo {S}oon? {C}hildren
  and {S}ocial {M}edia {A}pps,'' 2021, accessed on: Dec. 24, 2021. [Online].
  Available:
  \url{https://mottpoll.org/reports/sharing-too-soon-children-and-social-media-apps}
\BIBentrySTDinterwordspacing

\bibitem{papadakis19}
S.~Papadakis, N.~Zaranis, and M.~Kalogiannakis, ``Parental {I}nvolvement and
  {A}ttitudes {T}owards {Y}oung {G}reek {C}hildren's {M}obile {U}sage,''
  \emph{International Journal of Child-Computer Interaction}, vol.~22, no.
  100144, 2022.

\bibitem{papadakis22}
S.~Papadakis, F.~Alexandraki, and N.~Zaranis, ``Mobile {D}evice {U}se {A}mong
  {P}reschool-{A}ged {C}hildren in {G}reece,'' \emph{Education and Information
  Technologies}, vol.~27, pp. 2717--2750, 2022.

\bibitem{alqahtani16}
A.~M. Alqahtani, ``Keeping {S}afe {O}nline: {P}erceptionsof {G}ulf
  {A}dolescents,'' \emph{Journal of Education and e-Learning Research}, vol.~3,
  no.~4, pp. 150--155, 2016.

\bibitem{alqahtani17}
N.~Alqahtani, S.~Furnell, S.~Atkinson, and I.~Stengel, ``Internet {R}isks for
  {C}hildren: {P}arents' {P}erceptions and {A}ttitudes: an {I}nvestigative
  {S}tudy of the {S}audi {C}ontext,'' in \emph{Proceedings of Internet
  Technologies and Applications (ITA)}, 2017, pp. 98--103.

\bibitem{almogbel15}
A.~Almogbel, M.~Begg, and S.~Wilford, ``Analysis of the {R}elationship
  {B}etween {S}audi {A}rabia parents' {E}ducation and {E}conomic {L}evels and
  {P}arental {C}ontrol of {I}nternet {U}sage,'' \emph{The Macrotheme Review},
  vol.~4, no.~4, pp. 142--159, 2015.

\bibitem{hartikainen16}
H.~Hartikainen, N.~Iivari, and M.~Kinnula, ``Should {W}e {D}esign for
  {C}ontrol, {T}rust or {I}nvolvement? {A D}iscourses {S}urvey about
  {C}hildren's {O}nline {S}afety,'' in \emph{Proceedings of Interaction Design
  and Children (IDC)}, 2016, pp. 367--378.

\bibitem{ghosh18}
A.~K. Ghosh, K.~Badillo-Urquiola, S.~Guha, J.~J.~L. Jr., and P.~J. Wisniewski,
  ``Safety vs. {S}urveillance: {W}hat {C}hildren {H}ave to {S}ay about {M}obile
  {A}pps for {P}arental {C}ontrol,'' in \emph{Proceedings of CHI Conference on
  Human Factors in Computing Systems}, 2018, pp. 1--14.

\bibitem{qs21}
\BIBentryALTinterwordspacing
{QS Quacquarelli Symonds Limited}, ``King {A}bdulaziz {U}niversity ({KAU}),''
  2021, accessed on: Dec. 15, 2021. [Online]. Available:
  \url{https://www.topuniversities.com/universities/king-abdulaziz-university-kau}
\BIBentrySTDinterwordspacing

\bibitem{wikipedia20}
\BIBentryALTinterwordspacing
Wikipedia, ``Wikipedia -- {K}ing {A}bdulaziz {U}niversity,'' 2020, accessed on:
  Dec. 16, 2020. [Online]. Available:
  \url{https://en.wikipedia.org/wiki/King_Abdulaziz_University}
\BIBentrySTDinterwordspacing

\bibitem{educba20}
\BIBentryALTinterwordspacing
EDUCBA, ``Statistical {A}nalysis {T}ypes,'' 2020, accessed on: Apr. 2, 2021.
  [Online]. Available: \url{https://www.educba.com/statistical-analysis-types}
\BIBentrySTDinterwordspacing

\bibitem{campbell07}
I.~Campbell, ``Chi-{S}quared and {F}isher--{I}rwin {T}ests of {T}wo-by-{T}wo
  {T}ables with {S}mall {S}ample {R}ecommendations,'' \emph{Statistics in
  Medicine}, vol.~26, no.~19, pp. 3661--3675, 2007.

\bibitem{aldamegh14}
\BIBentryALTinterwordspacing
S.~Aldamegh, ``Sufficiency {L}ine in the {K}ingdom of {S}audi {A}rabia,'' 2014,
  accessed on: Dec. 16, 2020. [Online]. Available:
  \url{http://khair.ws/library/1899}
\BIBentrySTDinterwordspacing

\bibitem{worlddata2020}
\BIBentryALTinterwordspacing
{WorldData.info}, ``Inflation {R}ates in {S}audi {A}rabia,'' 2019, accessed on:
  Feb. 6, 2021. [Online]. Available:
  \url{https://www.worlddata.info/asia/saudi-arabia/inflation-rates.php}
\BIBentrySTDinterwordspacing

\bibitem{statista21}
\BIBentryALTinterwordspacing
{Statista}, ``Saudi {A}rabia {I}nflation {R}ate 2024,'' 2021, accessed on: Feb.
  6, 2021. [Online]. Available:
  \url{https://www.statista.com/statistics/268062/inflation-in-saudi-arabia}
\BIBentrySTDinterwordspacing

\bibitem{gas18b}
\BIBentryALTinterwordspacing
{General Authority for Statistics (GAS)}, ``Household {I}ncome and
  {E}xpenditure {S}urvey,'' 2018, accessed on: Feb. 6, 2021. [Online].
  Available: \url{https://www.stats.gov.sa/en/37}
\BIBentrySTDinterwordspacing

\bibitem{gas_edu_17}
\BIBentryALTinterwordspacing
------, ``Education and {T}raining,'' 2017, accessed on: Dec. 6, 2021.
  [Online]. Available: \url{https://www.stats.gov.sa/en/903}
\BIBentrySTDinterwordspacing

\bibitem{thomas19}
G.~Thomas, J.~A. Bennie, K.~D. Cocker, O.~Castro, and S.~J.~H. Biddle, ``A
  {D}escriptive {E}pidemiology of {S}creen-{B}ased {D}evices by {C}hildren and
  {A}dolescents: {A} {S}coping {R}eview of 130 {S}urveillance {S}tudies {S}ince
  2000,'' \emph{Child Indicators Research}, vol.~13, pp. 935--950, 2019.

\bibitem{sandercock13}
G.~R.~H. Sandercock and A.~A. Ogunleye, ``Independence of {P}hysical {A}ctivity
  and {S}creen {T}ime as {P}redictors of {C}ardiorespiratory {F}itness in
  {Y}outh,'' \emph{Pediatric Research}, vol.~73, no.~5, pp. 692--697, 2013.

\bibitem{aacp21}
\BIBentryALTinterwordspacing
{American Academy of Child and Adolescent Psychiatry}, ``Screen {T}ime and
  {C}hildren,'' 2021, accessed on: Jan. 9, 2021. [Online]. Available:
  \url{https://www.aacap.org/AACAP/Families_and_Youth/Facts_for_Families/FFF-Guide/Children-And-Watching-TV-054.aspx}
\BIBentrySTDinterwordspacing

\bibitem{barlow07}
S.~E. Barlow and {the Expert Committee}, ``Expert {C}ommittee {R}ecommendations
  {R}egarding the {P}revention, {A}ssessment, and {T}reatment of {C}hild and
  {A}dolescent {O}verweight and {O}besity: {S}ummary {R}eport,''
  \emph{Pediatrics}, vol. 120, pp. s164--s192, 2013.

\bibitem{przybylski17}
A.~K. Przybylski and N.~Weinstein, ``Digital {S}creen-{T}ime {L}imits and
  {Y}oung {C}hildren's {P}sychological {W}ell-{B}eing: {E}vidence {F}rom a
  {P}opulation-{B}ased {S}tudy,'' \emph{Child Development}, vol.~90, pp.
  e56--e65, 2019.

\bibitem{gplay20}
\BIBentryALTinterwordspacing
Google, ``Google {P}lay,'' 2020, accessed on: Dec. 18, 2020. [Online].
  Available: \url{https://play.google.com/store}
\BIBentrySTDinterwordspacing

\bibitem{xu11}
X.~Xu, Z.~M. Mao, and J.~A. Halderman, ``Internet {C}ensorship in {C}hina:
  {W}here {D}oes the {F}iltering {O}ccur?'' in \emph{Proceedings of Passive and
  Active Network Measurement (PAM)}, 2011, pp. 133--142.

\bibitem{citc21}
\BIBentryALTinterwordspacing
{Communications and Information Technology Commission (CITC)}, ``Web
  {F}iltering,'' 2021, accessed on: Dec. 21, 2021. [Online]. Available:
  \url{https://www.citc.gov.sa/en/Services/Pages/InternetFiltering.aspx}
\BIBentrySTDinterwordspacing

\bibitem{niaki21}
A.~A. Niaki, S.~Cho, Z.~Weinberg, N.~P. Hoang, A.~Razaghpanah, N.~Christin, and
  P.~Gill, ``I{CL}ab: {A} {G}lobal, {L}ongitudinal {I}nternet {C}ensorship
  {M}easurement {P}latform,'' in \emph{{IEEE} {S}ymposium on {S}ecurity and
  {P}rivacy ({S}\&{P})}, 2020, pp. 135--151.

\bibitem{google21}
\BIBentryALTinterwordspacing
Google, ``Apps \& {G}ames {C}ontent {R}atings on {G}oogle {P}lay,'' 2021,
  accessed on: Mar. 20, 2021. [Online]. Available:
  \url{https://support.google.com/googleplay/answer/6209544?p=appgame_ratings&visit_id=637518655539811089-2619076969&rd=1}
\BIBentrySTDinterwordspacing

\bibitem{boe21}
\BIBentryALTinterwordspacing
{Bureau of Experts at the Council of Ministers}, ``{P}ersonal {D}ata
  {P}rotection {R}egulation,'' 2021, accessed on: March 8, 2022. [Online].
  Available:
  \url{https://laws.boe.gov.sa/BoeLaws/Laws/LawDetails/b7cfae89-828e-4994-b167-adaa00e37188/1}
\BIBentrySTDinterwordspacing

\bibitem{alajlan21}
\BIBentryALTinterwordspacing
A.~AlAjlan, Z.~Y. Younes, Y.~Bugaighis, and K.~Blyth, ``Saudi {A}rabia:
  {P}ersonal {D}ata {P}rotection {L}aw {E}nacted,'' 2021, accessed on: Nov. 24,
  2021. [Online]. Available:
  \url{https://www.globalcompliancenews.com/2021/10/10/saudi-arabia-personal-data-protection-law-enacted-27092021/}
\BIBentrySTDinterwordspacing

\bibitem{gazette21}
\BIBentryALTinterwordspacing
{Saudi Gazette}, ``Saudi {A}rabia {A}pproves {N}ew law to {P}rotect {P}ersonal
  {D}ata,'' 2021, accessed on: Dec. 17, 2021. [Online]. Available:
  \url{https://saudigazette.com.sa/article/610907}
\BIBentrySTDinterwordspacing

\bibitem{sdaia_22}
\BIBentryALTinterwordspacing
{Saudi Data \& AI Authority (SDAIA)}, ``A {P}ress {R}elease by the {S}audi
  {D}ata \& {AI} {A}uthority {SDAIA} on the {P}ersonal {D}ata {P}rotection
  {L}aw in {S}audi {A}rabia,'' 2022, accessed on: Mar. 22, 2022. [Online].
  Available: \url{https://twitter.com/SDAIA_SA/status/1505140072655556614}
\BIBentrySTDinterwordspacing

\bibitem{sdaia22_2}
\BIBentryALTinterwordspacing
------, ``Statement from the {S}audi {D}ata \& {AI} {A}uthority {SDAIA}
  {R}egarding {P}ostponing the {E}nforcement of the {S}audi {P}ersonal {D}ata
  {P}rotection {L}aw,'' 2022, accessed on: Mar. 22, 2022. [Online]. Available:
  \url{https://twitter.com/SDAIA_SA/status/1506319750447644672}
\BIBentrySTDinterwordspacing

\bibitem{spa22}
\BIBentryALTinterwordspacing
{Saudi Press Agency (SPA)}, ``{SDAIA} {P}ostpones the {E}nforcement of the
  {S}audi {P}ersonal {D}ata {P}rotection {L}aw,'' 2022, accessed on: Mar. 22,
  2022. [Online]. Available:
  \url{https://www.spa.gov.sa/viewfullstory.php?lang=en&newsid=2339791#2339791}
\BIBentrySTDinterwordspacing

\bibitem{alghamdi21}
A.~K.~H. Alghamdi, S.~L.~T. McGregor, and W.~S. El-Hassan, ``Financial
  {L}iteracy, {S}tability, and {S}ecurity as {U}nderstood by {M}ale,''
  \emph{International Journal of Economics and Finance}, vol.~13, no.~7, pp.
  7--20, 2021.

\bibitem{almushare15}
\BIBentryALTinterwordspacing
K.~A. Almushare, ``Raising {F}inancial {A}wareness {A}mong {S}audi {Y}outh,''
  2015, accessed on: Dec. 23, 2021. [Online]. Available:
  \url{https://www.arabnews.com/economy/news/702496}
\BIBentrySTDinterwordspacing

\bibitem{kpmg20}
\BIBentryALTinterwordspacing
{KPMG}, ``Analysis of {H}ousehold {S}avings in {S}audi {A}rabia,'' 2021,
  accessed on: Dec. 23, 2020. [Online]. Available:
  \url{https://assets.kpmg/content/dam/kpmg/sa/pdf/2020/analysis-of-household-savings-in-saudi-arabia.pdf}
\BIBentrySTDinterwordspacing

\bibitem{baumrind66}
D.~Baumrind, ``Effects of {A}uthoritative {P}arental {C}ontrol on {C}hild
  {B}ehavior,'' \emph{Child Development}, vol.~37, no.~4, pp. 887--907, 1966.

\bibitem{dwairy06}
M.~Dwairy, M.~Achoui, R.~Abouserie, A.~Farah, A.~A. Sakhleh, M.~Fayad, and
  H.~K. Khan, ``Parenting {S}tyles in {A}rab {S}ocieties: {A} {F}irst
  {C}ross-{R}egional {R}esearch {S}tudy,'' \emph{Journal of Cross-Cultural
  Psychology}, vol.~37, no. 230, pp. 230--247, 2006.

\bibitem{qayyem21}
\BIBentryALTinterwordspacing
Values.sa, ``Qayyem,'' 2021, accessed on: Dec. 10, 2021. [Online]. Available:
  \url{https://values.sa/qayyem}
\BIBentrySTDinterwordspacing

\bibitem{misk21}
\BIBentryALTinterwordspacing
Misk, ``Misk {F}oundation,'' 2021, accessed on: Dec. 10, 2021. [Online].
  Available: \url{https://misk.org.sa/en}
\BIBentrySTDinterwordspacing

\bibitem{arabnews21}
\BIBentryALTinterwordspacing
{Arab News}, ``Misk {L}launches {V}ideo {G}ame {S}afety {P}latform in {S}audi
  {A}rabia,'' 2021, accessed on: Dec. 10, 2021. [Online]. Available:
  \url{https://www.arabnews.com/node/1604081/saudi-arabia}
\BIBentrySTDinterwordspacing

\bibitem{qayyem_faq21}
\BIBentryALTinterwordspacing
Values.sa, ``Qayyem - {FAQ}s,'' 2021, accessed on Dec. 17, 2021. [Online].
  Available: \url{https://values.sa/qayyem/en/faqs-en/}
\BIBentrySTDinterwordspacing

\bibitem{marvels21}
\BIBentryALTinterwordspacing
------, ``{M}arvels {A}vengers,'' 2021, accessed on: Dec. 12, 2021. [Online].
  Available: \url{https://values.sa/qayyem/games/marvels-avengers}
\BIBentrySTDinterwordspacing

\bibitem{zhao19}
J.~Zhao, G.~Wang, C.~Dally, P.~Slovak, J.~Edbrooke-Childs, M.~V. Kleek, and
  N.~Shadbolt, ``{'I make up a silly name'}: {U}nderstanding {C}hildren's
  {P}erception of {P}rivacy {R}isks {O}nline,'' in \emph{Proceedings of CHI
  Conference on Human Factors in Computing Systems}, 2019, pp. 1--13.

\bibitem{kennedy16}
L.~Zhang-Kennedy, C.~Mekhail, Y.~Abdelaziz, and S.~Chiasson, ``From {N}osy
  {L}ittle {B}rothers to {S}tranger-{D}anger: {C}hildren and {P}arents'
  {P}erception of {M}obile {T}hreats,'' in \emph{Proceedings of Interaction
  Design and Children (IDC)}, 2016, pp. 388--399.

\bibitem{kumar17}
P.~Kumar, S.~M. Naik, U.~R. Devkar, M.~Chetty, T.~L. Clegg, and J.~Vitak, ``No
  {T}elling {P}asscodes {O}ut {B}ecause {T}hey're {P}rivate': {U}nderstanding
  {C}hildren's {M}ental {M}odels of {P}rivacy and {S}ecurity {O}nline,''
  \emph{ACM Human-Computer Interaction}, vol.~1, no. CSCW, 2017.

\bibitem{dowthwaite20}
L.~Dowthwaite, H.~Creswick, V.~Portillo, J.~Zhao, M.~Patel, E.~P. Vallejos,
  A.~Koene, and M.~Jirotka, ``{"It's Your Private Information. It's Your
  Life."}: {Y}oung {P}eople's {V}iews of {P}ersonal {D}ata {U}se by {O}nline
  {T}echnologies,'' in \emph{Proceedings of Interaction Design and Children
  (IDC)}, 2020, pp. 121--134.

\bibitem{urquiola19}
K.~Badillo-Urquiola, D.~Smriti, B.~McNally, E.~Golub, E.~Bonsignore, and P.~J.
  Wisniewski, ``Stranger {D}anger! {S}ocial {M}edia {A}pp {F}eatures
  {C}o-{D}esigned with {C}hildren to {K}eep {T}hem {S}afe {O}nline,'' in
  \emph{Proceedings of Interaction Design and Children (IDC)}, 2019, pp.
  394--406.

\bibitem{urquiola17}
K.~Badillo-Urquiola, S.~Harpin, and P.~Wisniewski, ``Abandoned but {N}ot
  {F}orgotten: {P}roviding {A}ccess {W}hile {P}rotecting {F}oster {Y}outh from
  {O}nline {R}isks,'' in \emph{Proceedings of Interaction Design and Children
  (IDC)}, 2017, pp. 17--26.

\bibitem{fitton19}
D.~Fitton and J.~C. Read, ``Creating a {F}ramework to {S}upport the {C}ritical
  {C}onsideration of {D}ark {D}esign {A}spects in {F}ree-to-{P}lay {A}pps,'' in
  \emph{Proceedings of Interaction Design and Children (IDC)}, 2019, pp.
  407--418.

\bibitem{cavoukian11}
\BIBentryALTinterwordspacing
A.~Cavoukian, ``Privacy by {D}esign: {T}he 7 {F}oundational {P}rinciples,''
  2011, accessed on: Dec. 14, 2021. [Online]. Available:
  \url{https://www.ipc.on.ca/wp-content/uploads/Resources/7foundationalprinciples.pdf}
\BIBentrySTDinterwordspacing

\bibitem{yap20}
C.~E.~L. Yap and J.-J. Lee, ``{'Phone Apps Know a Lot about You!'}: {E}ducating
  {E}arly {A}dolescents about {I}nformational {P}rivacy through a {P}hygital
  {I}nteractive {B}ook,'' in \emph{Proceedings of Interaction Design and
  Children (IDC)}, 2020, pp. 49--62.

\bibitem{potapov20}
K.~Potapov and P.~Marshall, ``Life{M}osaic: {C}o-{D}esign of a {P}ersonal
  {I}nformatics {T}ool for {Y}outh,'' in \emph{Proceedings of Interaction
  Design and Children (IDC)}, 2020, pp. 519--531.

\bibitem{kumar18}
P.~Kumar, J.~Vitak, M.~Chetty, T.~L. Clegg, J.~Yang, B.~McNally, and
  E.~Bonsignore, ``Co-{D}esigning {O}nline {P}rivacy-{R}elated {G}ames and
  {S}tories with {C}hildren,'' in \emph{Proceedings of Interaction Design and
  Children (IDC)}, 2018, pp. 67--79.

\bibitem{gdpr21}
\BIBentryALTinterwordspacing
{Proton Technologies AG}, ``General {D}ata {P}rotection {R}egulation {GDPR}
  {C}ompliance {G}uidelines,'' 2021, accessed on: Apr. 13, 2021. [Online].
  Available: \url{https://gdpr.eu}
\BIBentrySTDinterwordspacing

\bibitem{ftc98}
\BIBentryALTinterwordspacing
{Federal Trade Commission (FTC)}, ``Children's {O}nline {P}rivacy {P}rotection
  {R}ule (\quotes{{COPPA}}),'' 1998, accessed on: Apr. 29, 2021. [Online].
  Available:
  \url{https://www.ftc.gov/enforcement/rules/rulemaking-regulatory-reform-proceedings/childrens-online-privacy-protection-rule}
\BIBentrySTDinterwordspacing

\bibitem{ico21}
\BIBentryALTinterwordspacing
{Information Commissioner's Office (ICO)}, ``Age {A}ppropriate {D}esign
  {C}ode,'' 2021, accessed on: Jan. 4, 2022. [Online]. Available:
  \url{https://ico.org.uk/for-organisations/guide-to-data-protection/ico-codes-of-practice/age-appropriate-design-a-code-of-practice-for-online-services}
\BIBentrySTDinterwordspacing

\bibitem{reyes18}
I.~Reyes, P.~Wijesekera, J.~Reardon, A.~E.~B. On, bbas Razaghpanah,
  NarseoVallina-Rodriguez, and S.~Egelman, ``\quotes{Won't {S}omebody {T}hink
  of the {C}hildren?} {E}xamining {COPPA} {C}ompliance at {S}cale,''
  \emph{Privacy Enhancing Technologies}, vol.~3, pp. 63--83, 2018.

\end{thebibliography}

\end{document}